\documentstyle[12pt]{article}

\newcommand{\be}{\begin{equation} }
\newcommand{\ee}{\end{equation} }
\newcommand{\ba}{\begin{array}}
\newcommand{\ea}{\end{array}}
\newcommand{\beq}{\begin{eqnarray}}
\newcommand{\eeq}{\end{eqnarray}}
\newcommand{\longeq}[2]{\displaystyle{ #1 }\\ \noalign{\vskip 1ex}
\displaystyle{#2}}
\newcommand{\llongeq}[3]{\displaystyle{ #1 }\\ \noalign{\vskip 1ex}
\displaystyle{#2} \\ \noalign{\vskip 1ex} \displaystyle{#3} }
\newcommand{\lllongeq}[4]{\displaystyle{ #1 }\\ \noalign{\vskip 1ex}
\displaystyle{#2} \\ \noalign{\vskip 1ex} \displaystyle{#3} \\ \noalign{\vskip
1ex}
\displaystyle{#4} }
\newcommand{\llllongeq}[5]{\displaystyle{ #1 }\\ \noalign{\vskip 1ex}
\displaystyle{#2} \\ \noalign{\vskip 1ex} \displaystyle{#3} \\ \noalign{\vskip
1ex}
\displaystyle{#4} \\ \noalign{\vskip 1ex} \displaystyle{#5} }
\newcommand{\lllllongeq}[6]{\displaystyle{ #1 }\\ \noalign{\vskip 1ex}
\displaystyle{#2} \\ \noalign{\vskip 1ex} \displaystyle{#3} \\ \noalign{\vskip
1ex}
\displaystyle{#4} \\ \noalign{\vskip 1ex} \displaystyle{#5} \\ \noalign{\vskip
1ex}
\displaystyle{#6}}
\newcommand{\lab} [1] {\label{eq:{#1}}}
\newcommand{\eqn} [1] {(\ref{eq:{#1}})}
\newcommand{\itGamma}{{\mit \Gamma}}

\newcommand{\itPsi}{{\mit \Psi}}

\newcommand{\itTheta}{{\mit \Theta}}
\newcommand{\itPi}{{\mit \Pi}}

\newcommand{\bm} [1] { { \mbox{\boldmath${#1}$}}  }
\newcommand{\bs}[1]{\!\! \mbox{\footnotesize{ {\boldmath {${#1}$}}}}}
\newcommand{\dsl}[2]{{#1}{\mbox{\hspace{-8pt}\hspace{#2}$\not$}
\hspace{8pt}\hspace{-#2}}}
\newcommand{\dslA}{\dsl{A}{0pt}}

\newcommand{\dslp}{\dsl{p}{1.5pt}}

\newcommand{\dslq}{\dsl{q}{.5pt}}

\newcommand{\dslk}{\dsl{k}{0pt}}
\newcommand{\dslL}{\dsl{L}{-.5pt}}
\newcommand{\dsll}{\dsl{l}{.5pt}}
\newcommand{\dslpart}{\dsl{\partial}{0pt}}
\newcommand{\dbox}{\,\framebox(7,7)[t]{}\,}
\newcommand{\Tprod}[1]{T\!\left({#1}\right)}

\newcommand{\KK}[1]{\left({#1}\right)}
\newcommand{\KKK}[1]{\left\{{#1}\right\}}

\newcommand{\kk}[1]{\mbox{{\small ({\normalsize${#1}$})}}}

\newcommand{\bra}[1]{\left\langle{#1}\right\vert}
\newcommand{\ket}[1]{\left\vert{#1}\right\rangle}
\newcommand{\commu}[2]{\bigl[\:{#1}\,,{#2}\,\,\bigr]}

\newcommand{\abs}[1]{\left\vert{#1}\right\vert}

\newcommand{\rmchar}[1]{\mbox{\rm{#1}}}

\newcommand{\ddd}[2]{\mathop{\partial_{{#1}{}}^{#2}}}
\newcommand{\der}[2]{\ddd{#1}{#2}\!}
\newcommand{\lrder}[2]{\ddd{#1}{#2}^{\leftrightarrow}\!}
\newcommand{\DDD}[2]{\mathop{D_{{#1}{}}^{#2}}}
\newcommand{\Der}[2]{\DDD{#1}{#2}\!}
\newcommand{\lrDer}[2]{\DDD{#1}{#2}^{\leftrightarrow}\!}

\newcommand{\inte}[1]{\int\!\!{#1}}
\newcommand{\integ}[3]{\int_{#1}^{#2}\!\!{#3}}

\newcommand{\bpsi}{\overline{\psi}}
\newcommand{\lrDerb}{\mathop{\dsl{D}{-1pt}}^{\leftrightarrow}\!}
\newcommand{\dslD}{\dsl{D}{-1pt}}
\newcommand{\lrderpartb}{\mathop{\dslpart}^{\leftrightarrow}\!}

\makeatletter
\@addtoreset{equation}{section}

\makeatother

\begin{document}
\title{Physical States and  Gauge Independence of \\ the Energy-Momentum Tensor
 \\
in Quantum Electrodynamics}
\author{Taro KASHIWA\thanks{e-mail:
taro1scp@mbox.nc.kyushu-u.ac.jp} and Naoki
TANIMURA\thanks{e-mail: tnmr1scp@mbox.nc.kyushu-u.ac.jp}\\
Department of Physics, Kyushu University\\ Fukuoka 812-81,
JAPAN\\\\}

%\date{\today}
\date{May 29, 1996}
\maketitle

\abstract{Discussions are made on the relationship between physical states
and gauge independence in QED. As the first candidate take the
LSZ-asymptotic states in a covariant canonical formalism to investigate
gauge independence of the (Belinfante's) symmetric energy-momentum
tensor. It is shown that expectation values of the energy-momentum tensor
in terms of those asymptotic states are gauge independent to all orders.
Second, consider gauge invariant operators of electron or photon , such as
the Dirac's electron or Steinmann's covariant approach, expecting a gauge
invariant result without any restriction. It is, however, demonstrated that
to single out gauge invariant quantities is merely synonymous to a gauge
fixing, resulting again in use of the asymptotic condition when proving
gauge independence. Nevertheless, it is commented that these invariant
approaches is helpful to understand the mechanism of the LSZ-mapping
and furthermore of quark confinement in QCD.  As the final candidate, it
is shown that gauge transformations are freely performed under the functional
representation or the path integral expression on account of the fact that the
functional
space is equivalent to a collection of infinitely many inequivalent Fock
spaces. The
covariant LSZ formalism is shortly reviewed and the basic facts on the
energy-momentum tensor are also illustrated.}

\newpage

\section{Introduction}
Symmetries play an important role in physics. In the usual
situation, invariance, once accepted, should be maintained
throughout the story. However gauge symmetry in quantum field
theory (QFT) has a rather different scenario: classically electric and
magnetic fields are gauge invariant but quantization has to be
carried out  in terms of gauge potentials by {\em fixing a gauge}
according to a standard recipe such as Dirac's \cite{DIRA} for
instance. Start with a classical Lagrangian, follow the canonical
procedure until setting up the Dirac bracket then utilize the
corresponding principle to obtain quantum theory. Accordingly
{\em each quantization in various gauges be carried out in a
different Fock space}, so what do we mean by ``gauge invariance''
of the S-matrix or expectation values of observables after quantization?
Moreover in order to get a representation in QFT, it is unavoidable
to introduce asymptotic fields, satisfying {\em linear} hyperbolic
field equations as well as being gauge invariant. Then how to bridge between
asymptotic fields and the Heisenberg fields (the LSZ-mapping) satisfying {\em
nonlinear} equation in a {\em fixed gauge}?

A way for proving gauge independence of, for instance, the S-matrix,  has so
far
been to show it under the perturbation theory with the use of a gauge dependent
photon propagator \cite{BLT}. There takes part a notion of physical states,
such as
the on-shell condition for electron or the photon polarization condition.  An
ambitious trial is to introduce gauge invariant fields for electron and photon
\cite{DT,ST,KT,LM}, expecting a fully gauge invariant result without any
conditions.
The approach furthermore leads us toward the understanding of the issue of
quark
confinement \cite{LMc}. Apart from these, the most widely adopted method of
proving gauge independence is that of functional integration \cite{AL}: start
with
some gauge by inserting the delta-function into path integral and
move to another gauge by means of the change of variables. However, as for the
authors' knowledge, there seems almost no clarification of this fact.

In this paper, we first study gauge independence of the energy-momentum tensor
in a covariant canonical formalism. Reasons for taking this issue are as
follows:
\begin{itemize}

\item (a) There have been trials for proving gauge independence of the
S-matrix \cite{BLT} but seems very few for checking that of
energy-momentum or the energy-momentum tensor. Needless to say that
energy-momentum must be gauge independent and so should be the
energy-momentum tensor coupled to gravity. Furthermore the
energy-momentum tensor is a good object to check some invariance; since it
is a composite operator so if higher order corrections are taken into account
there sometimes emerge serious problems which cannot be seen in a formal
discussion \cite{CCJ}: the self-stress problem of electron \cite{JRK} or the
trace
anomaly \cite{CJ} is well-known.  Freedman et al. \cite{FW}  studied that for
tadpole contributions of the energy-momentum tensor in the scalar QED but
they do not make any explicit calculation for other parts.

\item (b) It is preferable to utilize the covariant formulation under the
perturbation theory so that the covariant LSZ-formalism \cite{NN} must be
suitable.  It is then necessary
to impose physical state conditions and check whether those give us a gauge
independent result.
\end{itemize}

To make the situation clearer consider the canonical
energy-momentum tensor:
\be
T_{\mu\nu} \equiv   \sum_a {{\partial \cal L}\over {\partial
(\partial^\mu\phi^a
 )} }
\partial_\nu \phi^a  - g_{\mu\nu} {\cal L} .  \lab{I3}
\ee
In terms of the (classical) Lagrangian
\be
{\cal L}_c = \overline{\psi} \left({i\over2}\lrDerb  -m \right)
\psi - {1 \over 4} F_{\mu \nu} F^{\mu \nu},  \lab{I4}
\ee
with
\be
\ba{c}
\longeq{ \dslD \equiv  \gamma^\mu\! \Der{\mu}{} \ , \qquad
\Der{\mu}{} \equiv
\der{\mu}{} -ieA_\mu ,  }{ \Phi^* \lrDer{\mu}{} \Psi
\equiv \Phi^* \Der{\mu}{} \Psi-(\Der{\mu}{}\Phi)^* \Psi,}
\ea  \lab{I4a}
\ee
it reads
\be
T^c_{\mu\nu} \equiv \overline{\psi}{ i \over 2} \gamma_\mu
\lrder{\nu}{}
\psi -F_{\mu
\rho} \partial_\nu A^\rho -g_{\mu \nu} {\cal L}_c  \lab{I5}
\ee
which is apparently gauge variant, contrary to the Belinfante's
symmetric energy-momentum tensor,
\be
\itTheta^c_{\mu\nu} \equiv {i \over 4} \overline{\psi} \left(
\gamma_\mu
\lrDer{\nu}{}  + \gamma_\nu \lrDer{\mu}{} \right) \psi- F_{\mu
\rho}{F_\nu}^\rho -g_{\mu \nu} {\cal L}_c  \ , \lab{I6}
\ee
considered as the source of gravity, except for the scalar
case in which an improvement must be necessary \cite{CCJ}. However
there is no problem for energy-momentum: as is well-known from
the process of construction, difference is given by the total
derivative, $T^c_{\mu \nu} = \itTheta^c_{
\mu \nu} + \partial^\rho X_{[\rho, \mu] \nu}$ (with $\rho, \mu$
being antisymmetric), then
\be
 P_\mu = \int d^3 x \ T^c_{0 \mu} = \int d^3 x \ \itTheta^c_{0 \mu}.
\lab{I7}
\ee
Therefore energy-momentum itself and the Belinfante tensor is
gauge invariant and can be considered as observables classically.

The quantum Lagrangian is, due to
Nakanishi and Lautrup \cite{NL},
\be
{\cal L} \equiv {\cal L}_c +{\cal L}_{GF} =\overline{\psi}
\left({i\over2}\lrDerb  -m \right) \psi  - {1 \over 4} F_{\mu
\nu} F^{\mu \nu} - A^\mu \partial_\mu B + {
\alpha \over 2} {B}^2,
 \lab{I8}
\ee
with
\be
{\cal L}_{GF} \equiv - A^\mu \partial_\mu B + { \alpha \over 2}
{B}^2,
\ee
where $B$ is an auxiliary field, called the Nakanishi-Lautrup field,
and $\alpha$ is the gauge parameter. Although gauge has been
fixed we are left with the {\em BRS-symmetry} \cite{BRS,KO}:
\be
\ba{c}
\longeq{\delta_{B} A_\mu = \partial_\mu c,  \quad \delta_{B} c = 0 ,  \quad
\delta_{B} B = 0, \quad
\delta_{B} \overline{c} = -iB, }
{\delta_{B} \psi = i e c \psi, \quad \delta_{B} \overline{\psi} =- i e c
\overline{\psi} ,  }
\ea  \lab{a9}
\ee
with $c(\overline{c})$ being the Faddeev-Popov ghost (anti-ghost).
This keeps the following Lagrangian intact:
\be
{\cal L} + {\cal L}_{FP}= {\cal L}_c +{\cal L}_{GF} + {\cal L}_{FP} ,
\lab{Ilagrang}
\ee
with
\be
{\cal L}_{FP} \equiv i \partial_\mu
\overline{c}
\partial^\mu c.
\ee
Note that
\be
{\cal L}_{GF} + {\cal L}_{FP} = \delta_{B} \! \!\left( i {\alpha \over
2}
\overline{c} B - i \partial_\mu \overline{c} A^\mu \right).
\lab{Ilagbrs}
\ee
Thus gauge symmetry has been taken over by the BRS symmetry
in quantum theory.  The generator is called the BRS charge,
$Q_B$;
\be
\left[ A_\mu, Q_B\right] = i\delta_B A_\mu, \quad \left\{ \psi, Q_B
\right\} =i \delta_B \psi, \quad \dots   \lab{Icomm}
\ee
which gives a physical state condition:
\be
Q_B |  phys \rangle_{{}_B} = 0.  \lab{Iphys}
\ee
(Here $(\{a, b\}) [a, b]$ is the (anti-)commutator.)
Since the canonical energy-momentum tensor is not observable
even classically, we only concentrate on the Belinfante's one given
as
\be
\ba{l}
\longeq{\itTheta_{\mu \nu} = \itTheta^c_{\mu \nu} - \left( A_\mu
\partial_\nu B + A_\nu \partial_\mu B \right) + i \partial_\mu
\overline{c} \partial_\nu c +   i \partial_\nu
\overline{c} \partial_\mu c   -  g_{\mu \nu} \left( {\cal L}_{GF}+
{\cal L}_{FP} \right)}{ \hspace{5ex}= \itTheta^c_{\mu \nu} +
\delta_B \!\!\left[ -i
\partial_\mu
\overline{c} A_\nu   -i \partial_\nu \overline{c} A_\mu -g_{\mu
\nu} \left(  i {\alpha \over 2}
\overline{c} B - i \partial_\mu \overline{c} A^\mu\right) \right] ,}
\ea
\ee
which is, of course, BRS invariant;
\be
\Big[ Q_B , \itTheta_{\mu \nu} \Big] = 0 \ .
\ee
In view of  \eqn{Icomm} and \eqn{Iphys},
\be
{}_{{}_B} \langle  phys'  | \itTheta_{\mu \nu} |  phys \rangle_{{}_B}
= {}_{{}_B} \langle phys'  |
\itTheta^c_{\mu \nu} |  phys \rangle_{{}_B}\ .  \lab{Iexpect}
\ee
So when sandwiched
between {\em properly chosen  physical} states obeying
\eqn{Iphys}, the expectation value of the symmetric
energy-momentum tensor would become gauge independent
{\em provided that higher orders make no harmful effects.}

In order to check this, we work with the perturbative method in
the covariant formalism to calculate expectation values of the
energy-momentum tensor in terms of the loop expansion in \S2 and
convince ourselves of gauge independence (BRS invariance) to all orders. Main
machinery is the Ward-Takahashi relation with the aid of the
dimensional regularization which does not break the Poincar\'{e} as
well as the gauge invariance and is handier than the Pauli-Villars
regularization.

We then discuss about gauge invariant operators, expecting that the result
would
be gauge (BRS) invariant unconditionally. However, from the
discussion of \S3, we recognize that {\em picking up gauge invariant operators
for
basic fields, that is, for electrons  and photons, is merely synonymous to
gauge
fixing} so that we again need physical state conditions to prove gauge
independence.
We argue the LSZ-mapping in terms of these invariant fields also in \S3.

There is another physical state frequently adopted; the Gauss's law constraint
in the
$A_0=0$ gauge,
\be
\Phi({\bm x})| phys \rangle \equiv \left[ \sum_{k=1}^3 \left(
\partial_k  {\bm E}_k({\bm x}) \right) + J_0({\bm x}) \right] |phys
\rangle = 0,
\lab{Igauss}
\ee
where ${\bm E}_k( J_0)$ is the electric fields (the charge density).
According to the common sense in QFT \cite{PR} \eqn{Igauss}
implies $\Phi({\bm x})=0$, but there have been a number of
discussions in terms of this method. In \S4 we clarify the reason by
means of the functional representation. Also on account of that we can
build up the path integral formula starting from the Coulomb gauge
 and prove gauge independence more rigorously than the previous work
\cite{AL}. The final \S5 is devoted to discussion. In Appendix A, we review
the covariant LSZ formalism, and in Appendix B we study the violation of
the Ward-Takahashi identity for the energy-momentum tensor and then perform an
explicit renormalization for the energy-momentum tensor to illustrate that the
usual
procedure does indeed work well.

\section{LSZ and the Energy-Momentum Tensor}

In this section in order to examine gauge independence of the
expectation value we apply the covariant LSZ formalism to the
energy-momentum tensor and demonstrate that asymptotic states can
indeed be interpreted as physical states and there is no harmful
contribution from higher orders.

\subsection{Physical States in the LSZ formalism}
Start with a discussion on physical states in the LSZ formalism.
Details must be seen in Appendix A. In view of the BRS invariant
Lagrangian,
\be
{\cal L} +  {\cal L}_{FP}  =\overline{\psi} \left({i\over2}\lrDerb  -m
\right)
\psi  - {1 \over 4} F_{\mu
\nu} F^{\mu \nu}  - A^\mu \partial_\mu B + { \alpha \over 2} {B}^2
+ i \partial_\mu \overline{c} \partial^\mu c ,
  \lab{II1}
\ee
the ghosts are free, $\dbox c = \dbox \overline{c} =0$, all the time
so that we can reduce physical state $|  phys \rangle_{{}_B}$
\eqn{Iphys} to a simpler form: first decompose the total space such
that \cite{KO}
\begin{equation}
	{\cal V}\otimes\ket{0}_{FP}
\end{equation}
where $\ket{0}_{FP}$ is the vacuum of the FP ghost sector and ${\cal
V}$ is the remainder.
$Q_{B} |  phys \rangle_{{}_B} = 0$, \eqn{Iphys}, then implies
\begin{equation}
Q_{B} |  phys \rangle_{{}_B} =
Q_{B}\ket{phys}\otimes\ket{0}_{FP}=i\inte{d^{3}q}{\cal B}_{\bs q}
	\ket{phys}\otimes c^{\dag}_{\bs q}\ket{0}_{FP}=0  \ ,
\lab{B3}
\end{equation}
where the BRS charge has been given by
\begin{equation}
	Q_{B}=i\inte{d^{3}q}\left[c^{\dag}_{\bs q}{\cal B}_{\bs q}-
	{\cal B}_{\bs q}^{\dag}c_{\bs q}\right]
\end{equation}
with $c^{\dag}_{\bs q}(c_{\bs q})$ and ${\cal B}_{\bs
q}^{\dag}({\cal B}_{\bs q})$ being the creation (annihilation)
operator of the ghost and the $B$ field respectively. From
\eqn{B3} physical state in this reduced space reads
$ {\cal B}_{\bs q}\ket{phys}=0$ , giving \cite{NL}
\begin{equation}
      B^{(+)}\kk{x}\ket{phys}=0 \ .
\lab{psc}
\end{equation}
Thus throwing away the ghosts in \eqn{II1} we begin with
\be
{\cal L}  =\overline{\psi}
\left({i\over2}\lrDerb  -m \right) \psi  - {1 \over 4} F_{\mu
\nu} F^{\mu \nu} - A^\mu \partial_\mu B + {
\alpha \over 2} {B}^2.
\ee
Since $B(x)$ is free, $\dbox B(x) =0$, it goes to
the asymptotic field itself:${B}\kk{x}\longrightarrow  Z_3^{-1/2}
B^{\,as}\kk{x}$ ,
where ``as'' designates the asymptotic field (in or out),  so that the
physical state reads
\begin{equation}
      {B^{as}}^{(+)}\kk{x}\ket{phys;as}=0\ .
      \label{APSC}
\end{equation}
The commutation relations with respect to $B$,
\be
\commu{{A}_\mu\kk{x}}{{B}\kk{y}} =  -i\der{\mu}{x}D\kk{x-y} \ ,
\quad
\commu{{B}\kk{x}}{{B}\kk{y}} = 0 \ ,
\ee
trivially become
\be
\commu{ A^{as}_\mu\kk{x}}{{B^{as}}\kk{y}} =
-i\der{\mu}{x}D\kk{x-y} \ , \quad
\commu{{B^{as}}\kk{x}}{{B^{as}}\kk{y}} = 0 \ , \lab{IIasphot}
\ee
but those with electrons
\be
\commu{B(x)}{\psi(y)}=e\psi(y)D(x-y) \ ,\quad
\commu{B(x)}{\overline\psi(y)}=-e\overline\psi(y)D(x-y) \ , \lab{b11}
\ee
would become
\be
\commu{B^{as}(x)}{\psi^{as}(y)}=0  \ ,\quad
	\commu{B^{as}(x)}{\overline{\psi}^{as}(y) } =0  \ ;  \lab{IIaselec}
\ee
since the interaction should fade away in the asymptotic region. If we admit
\eqn{IIaselec} asymptotic states of electron would all be physical, that is,
${\psi^{as}(y)}$ is BRS invariant. As for those of photon, the $B$-state
\be
\ket{{\bm q};as} \equiv {{\cal
B}_{\bs q}^{as}}^{\dag} \ket{0}  \ ,
\ee
is physical owing to the second relation in \eqn{IIasphot}, but
\be
\ket{{\bm q}\sigma;as}\equiv {{\cal A}_{{\bs
q}\sigma}^{as}}^{\dag} \ket{0} \  ,
\lab{FS}
\ee
with ${{\cal A}_{{\bs q}\sigma}^{as}}^{\dag}\left({\cal A}_{{\bs
q}\sigma}^{as} \right)$ being the creation (annihilation) operator of
photon needs an  additional constraint to be physical.
Introduce the photon wave functions,
$\KKK{{h_{{\bs q}\sigma}}^{\!\!\mu}\kk{x}}$, $\KKK{{f_{{\bs
q}\sigma}}^{\!\!\mu}\kk{x}}$, defined through
\be
\dbox{h_{{\bs q}\sigma}}^{\!\!\mu}\kk{x}=
    {f_{{\bs q}\sigma}}^{\!\!\mu}\kk{x}\ ; \qquad   \dbox{f_{{\bs
        q}\sigma}}^{\!\!\mu}\kk{x} = 0 ;
\ee
those which are related each other:
\be
 {h_{{\bs q}\sigma}}^{\!\!\mu}\kk{x} =
     {1\over2}\left({\bm \nabla}^2\right)^{-1}\KKK{\KK{x_0\der{0}{}
     -{3\over2}} {f_{{\bs q}\sigma}}^{\!\!\mu}\kk{x}
     +g^{\mu 0}{f_{{\bs q}\sigma}}^{\!\!0}\kk{x} }\ .
 \lab{IIfandh}
\ee
Write the Fourier transformation as
\begin{equation}
     {f_{\bs q}}_\sigma^{\;\mu}\kk{x}=
     \inte{\frac{d^3p}{\sqrt{\kk{2\pi}^3 2p_0}}}\,
     {\xi_\sigma}^\mu\kk{\bm p}\,
     \varphi_{\bs q}\kk{\bm p}\,{\rmchar e}^{-ipx}\ ;
     \mbox{\qquad}p_0=\abs{\bm p}\ ,
\end{equation}
where $\varphi_{\bs q}\kk{\bm p}$'s are some orthonormal
set and ${\xi_\sigma}^\mu\kk{\bm p}$ is the photon polarization
vector satisfying
\begin{equation}
  {\xi_\sigma}^\mu\kk{\bm p}\,{\xi_\tau}_\mu\kk{\bm p} = \left(
\ba{cccc} 1 &  &  &\\
 & -1 & & \\ & & -1 & \\ & & & -1 \ea \right) \equiv \eta_{\sigma
\tau} \  ,
   \lab{IIortho}
\end{equation}
where the repeated indices imply a summation.
The physical state condition for \eqn{FS} then gives
\be
  {B^{as}}^{(+)}\kk{x}\ket{{{\bs q}\sigma};as}
      =\commu{{B^{as}}^{(+)}\kk{x}}
      {{{\cal A}_{{\bs q}\sigma}^{as}}^{\dag}}\ket{0}
      =  \partial_\mu {f_{{\bs q}\sigma}}^{\!\!\mu}\kk{x}\ket{0}\
\equiv {\overline f}_{{\bs q}\sigma}\kk{x}\ket{0}
\biggl(\longrightarrow 0\biggr) \ ,
\lab{IIasyphoton}
\ee
where use has been made of the first relation in \eqn{IIasphot}.
Thus
\be
 {\overline f}_{{\bs
q}\sigma}\kk{x} = \partial_\mu {f_{{\bs q}\sigma}}^{\!\!\mu}\kk{x} =  0  \ .
	\lab{PPC}
\ee
Note that the transversal condition of photon,
\be
{f_{{\bs q}\sigma}}^{\!\!0}\kk{x} =0,  \quad  \mbox{as well as}
\quad \sum_{l=1}^3
\nabla_l {f_{{\bs q}\sigma}}^{\!\!l}\kk{x} = 0 \ ,  \lab{IItransv}
\ee
belongs to \eqn{PPC}. In terms of the momentum representation, \eqn{PPC} turns
out to be
\be
p_\mu {\xi_\sigma}^\mu\kk{\bm p} = 0 \ . \lab{IIpsc}
\ee
Note that there remain only two components out of 16
${\xi_\sigma}^\mu\kk{\bm p}$'s; since there are
10 orthonormal conditions, \eqn{IIortho}, together with 4 physical
state conditions, \eqn{IIpsc}, leaving us two components. (Notations
and further details should be consulted for Appendix A.)

The LSZ reduction formula of some operator ${\cal O}$ for photon
reads, for example, as
\be
\ba{l}
 \llongeq{   \lefteqn{\bra{{\bm q}\sigma;out}
      \Tprod{{A}_{\nu}\kk{y}{\cal O}}\ket{0}}
      }
   { \hspace{ 10ex} = \inte{d^4x}
      \Bigl\{{{f_{{\bs q}\sigma}}^{\!\!\!\mu}}^{\ast}\kk{x}\dbox^x
      \bra{0}\Tprod{{A}_\mu\kk{x}{A}_{\nu}\kk{y}{\cal O}}\ket{0}}
    { \hspace{20ex} -\kk{1\!-\!{\alpha}}
      \bigm[{\overline h}_{{\bs q}\sigma}^{\;\ast}\kk{x}\dbox^x
      +{{f_{{\bs q}\sigma}}^{\!\!\!\mu}}^{\ast}\kk{x}\der{\mu}{x}\bigm]
      \bra{0}\Tprod{{B}\kk{x}{A}_{\nu}\kk{y}{\cal O}}
      \ket{0}\Bigr\} \ , }
\ea   \lab{IILSZ}
\ee
where ${\overline h}_{{\bs q}\sigma}^{\;\ast}\kk{x} \equiv
\partial_\mu {{h_{{\bs q}\sigma}}^{\!\!\mu}}^{\ast}\kk{x}$.
Imposing the physical state condition
\eqn{PPC} to \eqn{IILSZ} by the help of  \eqn{IIfandh} and
\eqn{IItransv} we find that the $B$-containing term in the right
hand side is dropped, giving a na\"{\i}ve amplitude consisting solely
of $A_\mu$'s. The reduction for electrons can be obtained in a usual manner.
Accordingly the task is to calculate the vacuum expectation value of the
energy-momentum tensor,
\be
\itTheta^{\mu\nu} \equiv
\itTheta^{\mu\nu}_{\rm g}+\itTheta^{\mu\nu}_{\rm m} ,  \lab{b24}
\ee
where
\be
\itTheta^{\mu\nu}_{\rm g}\equiv -F^{\mu\rho}F^{\nu}_{\ \rho}
-(A^\mu
\der{}{\nu}B  + A^\nu \der{}{\mu}B) -g^{\mu\nu} \left[  - {1 \over
4} F_{\mu
\nu} F^{\mu \nu} - A^\mu \partial_\mu B + { \alpha \over 2} {B}^2
\right] ,   \lab{emt}
\ee
is the photon part and
\be
\itTheta^{\mu\nu}_{\rm m}\equiv {1\over4}\bpsi
i(\gamma^\mu\!\lrDer{}{\nu} + \gamma^\nu\!\lrDer{}{\mu})\psi
-g^{\mu\nu} \overline{\psi}
\left({i\over2}\lrDerb  -m \right) \psi  \  ,
\ee
is the (gauge invariant) electron part. Remove external legs from the
vacuum expectation value and multiply photon wave functions, ${{f_{{\bs
q}\sigma}}^{\!\!\!\lambda}}\kk{x}$'s, or the free spinors, $\overline{u}({\bm
k'}),
u({\bm k})$, to obtain the desired quantity;  energy-momentum tensor in the
physical
state.

It is convenient to introduce the generating
functional such that
\be
\ba{l}
\longeq{
\exp\left[iW[J_{\mu},J,\overline\eta,\eta,\tau_{\mu\nu}^{\rm m}
		,\tau_{\mu\nu}^{\rm g}]\right] \equiv \exp\Big[iW[{\bm J}
, {\bm \eta} , {\bm \tau}]  \Big] }
 { \equiv \bra{0} T^* \exp\left[i\int\!\!d^{4}x\left\{ J_{\mu}A^{\mu}+JB
	+\overline\eta\psi+\bpsi\eta+\tau_{\mu\nu}^{\rm m}\itTheta^{\mu\nu}_{\rm m}
	+\tau_{\mu\nu}^{\rm g}\itTheta^{\mu\nu}_{\rm
g}\right\}\right]  \ket{0} }
\ea	  \lab{B1}
\ee
from which we obtain the connected Green's
functions,
\be
\ba{l}
\llllongeq
{ G^{\mu \nu; \lambda_1 \cdots
\lambda_m}_{\rm a (= g \, or \, m)}(0 ; x_1, \cdots,  x_n; y_1, \cdots y_m ,
z_1,
\cdots, z_m)  }
{ \equiv    \Biggl[{\delta \over   \delta \tau_{\mu\nu}^{\rm a}(0) } \Biggr]
\Biggl[{
\delta^n \over   i\delta J_{\lambda_1}(x_1) \cdots  i \delta J_{\lambda_n}(x_n)
} \Biggr] {
\delta^{2m} W[{\bm J} , {\bm \eta} , {\bm \tau}]  \over   \delta
\overline{\eta}(y_1)
\cdots
\delta \overline{\eta}(y_m) \delta \eta(z_1) \cdots
 \delta \eta(z_m)
 }   \  , }
{ =
\bra{0} T^* \itTheta^{\mu\nu}_{\rm a}(0)  A_{\lambda_1}(x_1) \cdots
A_{\lambda_n}(x_n) \psi(y_1)  \cdots \psi(y_m)
\overline{\psi}(z_1) \cdots \overline{\psi}(z_m) \ket{0}_{\rm conn} }
{ \hspace{6.5ex} \equiv  \int\!\! \prod_{j=1}^n {d^4
q_j \over(2\pi)^{4}} \prod_{l=1}^m {d^4 k_l \over(2\pi)^{4}}{d^4
p_l\over(2\pi)^{4}}
\  \exp\left[ -i
\sum_{j=1}^n q_j x_j - i \sum_{l=1}^m \left( k_l y_l-p_l z_l \right)
\right] \   } { \hspace{28ex} \times G^{\mu \nu \lambda_1 \cdots
\lambda_n}_{\rm a}(q_1, \cdots ,q_n ; k_1, \cdots,  k_m,  p_1,
\cdots, p_m) \ , }
 \ea    \lab{IIphself}
\ee
where $T^*$ designates a covariant $T$-product. Calculations are performed by
means of the
loop expansion.

Our ingredients are summarized as follows:
\begin{itemize}

\item (i) The physical state conditions: for photon
\be
q_\mu {\xi_\sigma}^\mu\kk{\bm q} = 0 \  .  \lab{pscB}
\ee
For electron
\be
\ba{l}
\longeq{ \left(\dslp -m\right)u(\bm{p},s) =0 \ , }
{ \overline{u}(\bm{k},s)\left(\dslk -m\right) =0  \ . }
\ea     \lab{IIfrelec}
\ee

\item (ii) Dimensional regularization; which preserves both gauge
symmetry and the Poincar\'{e} symmetry. (Note that using a na\"{\i}ve cutoff
breaks the situation; see Appendix B.)

\item (iii) The notion of finiteness of the energy-momentum tensor
\cite{CCJ,FW};
under which we can proceed only with the unrenormalized form. Divergences can
be subtracted in a gauge invariant way in anytime. (A short discussion on
renormalization is
seen also in Appendix B.)

\end{itemize}

\subsection{Tree Calculation}
The tree graphs give us basic vertices of the Feynman
rule:

\input epsf.tex
%\leavevmode

\be
\ba{l}
\llongeq{ \hspace{-5ex} \raisebox{-5ex}{ \epsfbox{Fig1a.epsf}} : \quad  {G^{\mu
\nu;
\lambda
\kappa}_{\rm g}(q, q')}^{(0)}
\equiv { -i \over q^2}X^{\mu\nu\lambda\kappa}(q,q') {-i\over q'^2} \  , }
{ \raisebox{-4ex}{\epsfbox{Fig1b.epsf}} : \quad   { G^{\mu \nu }_{\rm m}( k,
p)}^{(0)}
\equiv { i \over \dslk - m}Y^{\mu\nu}(k+p){ i \over\dslp - m} \  , }
{\raisebox{-12ex}{\epsfbox{Fig1c.epsf}} : \quad  { G^{\mu \nu;\lambda }_{\rm
m}(q,
k, p)}^{(0)}_{\rm 1PI}
\equiv { i \over \dslk -m}{-i e \over q^2} Z^{\mu\nu\lambda}(q)
{ i \over \dslp - m} \  , }
\ea   \lab{IIfeynmrule}
\ee
where the cross, ``$\times$'' , and ``1PI''  designate the insertion of
$\itTheta_{\mu \nu}$ and
the one-particle-irreducible part respectively.
$X^{\mu\nu\lambda\kappa}(q,q')$  in
\eqn{IIfeynmrule} is explicitly given as
\be
 X^{\mu\nu\lambda \kappa}(q,q') \equiv \widetilde X^{\mu\nu\lambda
\kappa}(q,q') - \left( q^\lambda X^{\mu\nu\kappa}(q,q') + {q'}^\kappa
X^{\mu\nu\lambda}(q,q') \right)   + \alpha q^\lambda {q'}^\kappa g^{\mu\nu}\ ,
\lab{Xes}
\ee
where
\be
 \widetilde X^{\mu\nu\lambda \kappa}(q,q') \equiv
(\delta^{\mu}_{\rho}\delta^{\nu}_{\sigma}
+\delta^{\mu}_{\sigma}
\delta^{\nu}_{\rho}-{1\over2}g^{\mu\nu}g_{\rho\sigma})
(q^{\rho}g^{\lambda\alpha}-q^{\alpha}g^{\lambda\rho})
({q'}^{\sigma}
\delta^{\kappa}_{\alpha}- q'_{\alpha} g^{\kappa\sigma})\ ,   \lab{IIXa}
\ee
comes from $FF$ term in \eqn{emt} thus is gauge invariant but the remaining
terms
are from $AB$ and $BB$ and then gauge variant:
\be
	X^{\mu\nu\kappa}(q,q')\equiv  (\delta^{\mu}_{\rho}\delta^{\nu}_{\sigma}
	+\delta^{\mu}_{\sigma}\delta^{\nu}_{\rho}-g^{\mu\nu}g_{\rho\sigma}) q^{\rho}
	d^{\kappa\sigma}(q') \ , \lab{IIXb}
\ee
where
\be
d^{\mu\nu}(q) \equiv  g^{\mu\nu}-(1-\alpha){{q}^\mu {q}^\nu \over
{q}^2} \  ,
\lab{IInpropa}
\ee
is the numerator of the photon propagator,
\be
D^{\mu \nu}(q) \equiv  -i  {d^{\mu \nu}(q) \over q^2} \ .
\ee
Note that the transverseness of $\widetilde
X^{\mu\nu\lambda
\kappa}(q,q')$,
\begin{equation}
q_\lambda\widetilde X^{\mu\nu\lambda \kappa}(q,q')= q'_\kappa
\widetilde X^{\mu\nu\lambda \kappa}(q,q')\ =0 \  ,
\end{equation}
and the structure of gauge dependent terms: those depend on the external
photon momentum as well as the index, i.e. , on $q^\lambda$, and/or
$q'^\kappa$, leaving no effect owing to the
physical photon condition \eqn{pscB}.  {\em ${ G^{\mu
\nu; \lambda \kappa}_{\rm g}(q,q')}^{(0)}
$ is therefore gauge independent.}

Next check ${ G^{\mu \nu }_{\rm m}(k, p)}^{(0)}$: $Y^{\mu\nu}(q)$ in
\eqn{IIfeynmrule} is given by
\be
Y^{\mu\nu}(q) \equiv {1\over2}\itGamma^{\mu\nu\lambda
}q_\lambda+mg^{\mu\nu}
\ ,
\lab{IIelect}
\ee
with
\be
\itGamma^{\mu\nu\lambda } \equiv  {1\over2}(\gamma^\mu
g^{\nu\lambda}+\gamma^\nu g^{\mu\lambda})-\gamma^\lambda g^{\mu\nu} \
.    \lab{IIgamma}
\ee
According to our assumption, \eqn{b11} to \eqn{IIaselec},
there is no gauge dependence in the electron
sector: {\em
$Y^{\mu\nu}$ is gauge invariant so is ${ G^{\mu \nu }_{\rm m}( k, p
)}^{(0)}$.}

The 1PI part of ${  G^{\mu \nu ;\lambda }_{\rm m}(q; k, p)}^{(0)}$,
$Z^{\mu\nu\lambda}(q)$ in \eqn{IIfeynmrule}, is expressed, in terms of
\eqn{IIgamma}, by
\be
Z^{\mu\nu\lambda}(q) \equiv \itGamma^{\mu\nu\rho}d_\rho^{\
\lambda}(q) \ .  \lab{IIvertex}
\ee
As is seen from Fig.2  reducible graphs take part in
$$ \epsfbox{Fig2.epsf} $$
giving totally
\be
\ba{l}
\llongeq{{\dslk-m\over i}{G^{\mu\nu;\lambda}_{\rm m}(q; k,
p)}^{(0)}{\dslp-m\over i}}
    {={ie\over q^2}\left[Y^{\mu\nu}(k\!+\!p\!-\!q){1\over \dslp -\dslq -m}
    \gamma^{\lambda}
    +\gamma^{\lambda}{1\over \dslk + \dslq -m}Y^{\mu\nu}(k\!+\!p\!+\!q)
    -\itGamma^{\mu\nu\lambda}\right] }{
	-{ie(1-\alpha) q^{\lambda}\over(q^2)^{2}}\!\left[Y^{\mu\nu}(k\!+\!p\!-\!q)
	{1\over\dslp-\dslq-m}(\dslp-m)-(\dslk-m)
	{1\over\dslk+\dslq-m}Y^{\mu\nu}(k\!+\!p\!+\!q)\right], }
	\ea
\ee
whose $\alpha$-dependent terms vanish due to the factor $q^\lambda$ or
$(\dslk-m)$
as well as $(\dslp-m)$.

Finally ${G^{\mu\nu;\lambda}_{\rm g}(q; k,
p)}^{(0)}$ is also gauge invariant:
\be
\ba{l}
	\llongeq{{\dslk-m\over i}{G^{\mu\nu;\lambda}_{\rm g}(q; k,
p)}^{(0)}{\dslp-m\over i}}
	{= {\mbox{\normalsize$-ie$}\over\mbox{\normalsize$q^{2}(k-p)^{2}$}}
	\gamma^{\rho}
	\widetilde X^{\mu\nu\lambda}_{\ \ \ \rho}(q,k\!-\!p) }
    {  +{\mbox{\normalsize$ie$}\over\mbox{\normalsize$q^{2}(k-p)^{2}$}}
    \left\{q^{\lambda}\gamma^{\rho}X^{\mu\nu}_{\ \ \rho}(q,k\!-\!p)+
	(\dslk-\dslp)\left[X^{\mu\nu\lambda}(k\!-\!p,q)-\alpha
	q^{\lambda}g^{\mu\nu} \right]\right\} \ ;  }
\ea
\ee
since the second term in the right hand side again vanishes because of
$q^\lambda$ and $(\dslk-m)$ as well as $(\dslp-m)$. (Recall that $\widetilde
X^{\mu\nu\lambda \rho}(q, q')$ is gauge invariant.)

All the tree graphs coupled to physical states are thus gauge independent.

\subsection{One-loop Calculation}
\paragraph{${ G^{\mu \nu; \lambda \kappa}_{\rm g}(q,q')}^{(1)}$}:
contributions are from the graphs, $p_1$ and $p_2$.
$$ \epsfbox{Fig3.epsf} $$
\be
\ba{l}
\longeq{ { G^{\mu \nu ;\lambda \kappa}_{\rm g}(q,q')}^{(1)}}{\hspace{5ex}
= { -i \over q^2} X^{\mu\nu\lambda \rho }(q,q')
{ -i \over q'^2} \itPi_{\rho\sigma}(q'){ -i \over q'^2}d^{\sigma\kappa}(q')
+ { -i \over q^2} d^{\lambda\rho}(q) \itPi_{\rho\sigma}(q)
{ -i \over q^2}X^{\mu \nu\sigma \kappa }(q,q')  { -i \over q'^2} \  ,}
\ea  \lab{B44}
\ee
where $ \itPi^{\rho\sigma}(q)$ is the vacuum polarization,
\be
\ba{l}
\longeq{  \itPi^{\rho\sigma}(q)\equiv -e^{2}\int\!\!{d^{n}l\over(2\pi)^{n}}
	tr\left[\gamma^{\rho}{1\over\dsll+\dslq-m}\gamma^{\sigma}
	{1\over\dsll-m}\right]  \ ; \quad \itPi^{\rho\sigma}(q)
      =(q^{2}g^{\rho\sigma}-q^{\rho}q^{\sigma})\itPi(q^{2})\ , }{
      \itPi(q) \equiv -ie^{2}{2\,{\rmchar{tr}}\mbox{\bf1}\over(4\pi)^{2}}
	\itGamma(2-{n\over2})\int^{1}_{0}\hspace{-8pt}dx\ x(1-x)
	\left(m^{2}-x(1-x)q^{2}\over4\pi\right)^{{n\over2}-2} \ ; }
\ea	\lab{IIvacpol}
\ee
obeying the transversal condition:
\be
q_{\rho}\itPi^{\rho\sigma}(q)=0 \ .   \lab{tr}
\ee
As was mentioned before, gauge dependent parts, in $X^{\mu \nu \lambda
\kappa}$ \eqn{Xes} and
$d^{\lambda \kappa}$ \eqn{IInpropa}, are proportional to $q^\lambda$ and/or
$q'^\kappa$, then they vanish on account of the physical photon condition
\eqn{pscB} when the momentum is external or of the transverseness of the vacuum
polarization
\eqn{tr} when it is internal.

\paragraph{${ G^{\mu\nu; \lambda \kappa}_{\rm m}(q, q')}^{(1)}$}: graphs are
given in $p_3 \sim p_6$.
\be
{ G^{\mu \nu; \lambda \kappa}_{\rm m}(q, q')}^{(1)} \equiv { -i \over q^2}
{d^\lambda}_\rho(q) \itPi^{\mu\nu;\rho\sigma}(q, q'){d_\sigma}^\kappa(q') { -i
\over
q'^2} \  ,
\lab{II1PR}
\ee
where
\be
\ba{l}
\lllongeq{ \itPi^{\mu\nu;\lambda\kappa}(q, q') \equiv
ie^{2}\inte{d^{n}l\over(2\pi)^{n}} {\rm tr}
\Bigg[\itGamma^{\mu\nu\lambda}
	 {1 \over \dsll - {\dslq' / 2}-m} \gamma^{\kappa}
	 {1\over \dsll+{\dslq' / 2}-m} }
   {\hspace{28ex}    +  \gamma^{\lambda}{1\over\dsll+{\dslq / 2}-m}
     \itGamma^{\mu\nu\kappa}{1\over\dsll-{\dslq / 2}-m}  }
{ \hspace{8ex}  -\gamma^{\lambda}{1\over\dsll+{\dslq / 2}+{\dslq' / 2}-m}
	 Y^{\mu\nu}(2l){1\over\dsll-{\dslq / 2}-{\dslq' / 2}-m}
	 \gamma^{\kappa}{1\over\dsll-{\dslq / 2}+{\dslq' / 2}-m} }
	{ \hspace{8ex}
-\gamma^{\lambda}{1\over\dsll+{\dslq/ 2}-{\dslq' / 2}-m}
	 \gamma^{\kappa}{1\over\dsll+{\dslq/ 2}+{\dslq' / 2}-m}Y^{\mu\nu}(2l)
	 {1\over\dsll-{\dslq/ 2}-{\dslq' / 2}-m}\Bigg] \ ,  }
\ea  \lab{B47}
\ee
which is free from gauge dependence, reflecting the gauge independence of
$\itTheta^{\mu\nu}_{\rm
m}$. The only dangerous part is therefore the gauge term in the propagators,
${d^\lambda}_\rho(q), {d_\sigma}^\kappa(q')$. However,
$\itPi^{\mu\nu;\lambda\kappa}(q,q')$ has a remarkable property,
\begin{equation}
    q_{\lambda}\itPi^{\mu\nu;\lambda\kappa}(q,q')
    =q'_{\kappa}\itPi^{\mu\nu;\lambda\kappa}(q,q')=0 \ ,
	\lab{trv}
\end{equation}
with which there remains no gauge dependence in \eqn{II1PR}. (Or the physical
state
condition for photon \eqn{pscB} also wipes out the gauge dependence in this
case.)
The relation
\eqn{trv} is easily recognized by means of a simple and straightforward
manipulation such that
$\dslq = (\dsll +\dslq/2-m) - (\dsll - \dslq/2 -m)$ or the Ward-Takahashi
relation
discussed below.

\paragraph{${G^{\mu\nu}_{\rm m}(k,p)}^{(1)}$}: graphs are $f_1 \sim f_5$.
$$ \epsfbox{Fig4.epsf} $$
\be
\ba{l}
 \llllongeq{ {\dslk-m\over i}{G^{\mu\nu}_{\rm m}(k,p)}^{(1)}
    {\dslp-m\over i} }
{= -ie^{2}\inte{{d^{n}l\over(2\pi)^{n}}\Big\{{1\over l^{2}}
	\bigg[-\itGamma^{\mu\nu\rho}{1\over
	\dslp -\dsll -m}\gamma_{\rho}-\gamma_{\rho}{1\over \dslk
	-\dsll -m}\itGamma^{\mu\nu\rho}} }
	 { \ \ \  +Y^{\mu\nu}(k+p){1\over \dslp -m}\gamma^{\rho}{1\over
\dslp -\dsll
	-m}\gamma_{\rho}+\gamma_{\rho}{1\over \dslk -\dsll
	 -m}\gamma^{\rho}{1\over \dslp -\dsll
-m}Y^{\mu\nu}(k+p) }
	{ \ \ \ +\gamma^{\rho}{1\over \dslk -\dsll
-m}Y^{\mu\nu}(k+p-2l){1\over \dslp
	-\dsll -m}\gamma_{\rho}\bigg]  }
{  \ \  \ -{(1-\alpha)\over (l^{2})^{2}}\left[(\dslk -m){1\over \dslk -\dsll
	-m}Y^{\mu\nu}(k+p-2l){1\over \dslp -\dsll -m}(\dslp -m)
	-Y^{\mu\nu}(k+p)\right]\Big\}  \ . }
\ea
\ee
The gauge dependence appears only in the last line whose first term, however,
vanishes by the on-shell condition of electron \eqn{IIfrelec} so does the
second
term owing to the property of the dimensional  regularization:
\be
	\int {{d^{n}l\over(2\pi)^{n}}}{1\over (l^{2})^N} =0 \ ;  \qquad \mbox{$N$:
integer}.
	\lab{IIint}
\ee

\paragraph{${G^{\mu\nu}_{\rm g}(k,p)}^{(1)}$}: there contributes only one
graph,
$f_6$.
\be
\ba{l}
\lllllongeq{ {\dslk-m\over i}{G^{\mu\nu}_{\rm g}(k,p)}^{(1)} {\dslp-m\over i} }
  {=  ie^{2}\inte{{d^{n}l\over(2\pi)^{n}}}
    {1\over L^{2}(L-Q)^{2}}
    X^{\mu\nu\lambda\kappa}(Q-L,L)\gamma_{\lambda}
    {1\over \dslp +\dslL -m}\gamma_{\kappa}  }
  { =  ie^{2}\inte{{d^{n}l\over(2\pi)^{n}}}
	{1\over L^{2}(L-Q)^{2}}
	\left[\widetilde X^{\mu\nu\lambda\kappa}(Q-L,L)\gamma_{\lambda}
	{1\over \dslp +\dslL -m}\gamma_{\kappa} \right.  }
   {  - X^{\mu\nu\kappa}(Q-L,L)(\dslk -m) {1\over \dslp +\dslL -m}
	\gamma_{\kappa}  }
	{ +X^{\mu\nu\lambda}(L,Q-L)\gamma_{\lambda}
	{1\over \dslp +\dslL -m}(\dslp -m)   }
{    \left.	+\alpha g^{\mu\nu}
	\left\{{{\dslk +\dslp}\over2}-m-(\dslk -m){1\over \dslp +\dslL -m}
	(\dslp -m)\right\}\right]  \ ,   }
\ea
\ee
where
\be
    L\equiv l+{Q\over2}\ , \qquad \qquad Q\equiv k-p\ .
\ee
In the final expression, terms from the second to the last line are gauge
dependent but the on-shell condition of electron wipes them out.

Note that the important relations for obtaining the gauge independent results
are \eqn{tr}
and \eqn{trv}. We must discuss
${G^{\mu \nu ; \lambda}_{\rm a}(q; k, p)}^{(1)}$ to complete the one loop
calculation.
The scenario, however, can be realized and furthermore generalized to any order
with
the aid of the Ward-Takahashi relation.

\subsection{General Proof for Gauge Independence}

We here show that gauge independence of the energy-momentum tensor
holds in any order of the loop expansion. However as is seen in the
following, {\em all photon lines can be treated as external} and {\em there
needs only for considering the tree and the one loop graphs of electron}.
To grasp this we should note that
\begin{itemize}

\item (a) All gauge dependent terms, in
$\itTheta^{\mu\nu}_{\rm g}$ (, in view of
$X^{\mu\nu\lambda \kappa}(q,q')$ \eqn{Xes},) and the
propagator \eqn{IInpropa}, possess a momentum contractible with a
vertex or a photon wave function. The latter vanishes trivially on
account of the physical photon condition
\eqn{pscB} so that we concentrate on the former. To
check gauge independence is therefore to check the consequence of
the photon momentum contraction to the vertex.  The procedure is exactly
the same as studying gauge independence of the S-matrix; in other words,
checking
the cancellation of gauge terms in the photon propagator \cite{BD,BLT}.

\item (b) On the other hand, the insertion of $\itTheta^{\mu\nu}_{\rm m}$
forms new types of graphs.  However $\itTheta^{\mu\nu}_{\rm m}$ itself is
gauge invariant so that gauge dependence lies in the photon
propagator. Checking gauge independence is again realized by the same
manipulation of momentum as in (a) to those new graphs.

\item(c) Because of the fact (a) any internal photon line can be cut
out. Graphs that must be
taken into account are therefore the tree and the one loop graphs of
electron.

\end{itemize}
The program is carried out by means of the Ward-Takahashi
relation (WT) derived by
applying the BRS transformation to the
generating functional $W$ \eqn{B1} \cite{KO}:
\be
\bra{0} \left[ Q_B , T^*\exp\left[i\int\!\!d^{4}x\left\{ J_{\mu}A^{\mu}+JB
	+\overline\eta\psi+\bpsi\eta+\tau_{\mu\nu}^{\rm m}\itTheta^{\mu\nu}_{\rm m}
	+\tau_{\mu\nu}^{\rm g}\itTheta^{\mu\nu}_{\rm
g}\right\}\right]  \right] \ket{0} = 0 \ ,
\ee
which becomes in view of \eqn{Icomm} and \eqn{a9}
\be
 \Biggl[e \left( \overline{\eta} { \delta \over \delta\overline{\eta} }
	-\eta{\delta \over \delta \eta} \right)
	+ \dbox {\delta\over i\delta J}
	+(\delta^{\mu}_{\rho}\delta^{\nu}_{\sigma}
	+\delta^{\mu}_{\sigma}\delta^{\nu}_{\rho}- g^{\mu\nu} g_{\rho\sigma} )
	\partial^{\rho}\!\! \left( \! \tau_{\mu\nu}^{\rm g}\partial^{\sigma}
	{\delta\over i\delta J} \! \right)\Biggr]W[{\bm J},{\bm\eta},{\bm\tau}]
	+ \  i \partial^\rho J_{\rho}=0 \ .
    \lab{GWT}
\ee
In order to simplify a discussion we further introduce the generating
functional of photon amputated Green's functions,
\be
\itGamma[{\bm A};{\bm\eta},{\bm\tau}]
\equiv W[{\bm J},{\bm\eta},{\bm\tau}]
-\int d^4 x J^{\mu}(x){\cal A}_{\mu}(x)-\int d^4 x J(x){\cal B}(x)  \ ,
\ee
where
\be
{\delta W[{\bm J},{\bm\eta},{\bm\tau}]\over\delta J^{\mu}}
\equiv {\cal A}_{\mu}\ ,
\hspace{30pt}{\delta W[{\bm J},{\bm\eta},{\bm\tau}]\over \delta J}
\equiv {\cal B}\ .
\ee
$\itGamma$ also obeys WT obtaining from \eqn{GWT}:
\be
\ba{l}
\longeq{\Biggl[e \left( \overline{\eta} { \delta \over \delta\overline{\eta} }
	-\eta{\delta \over \delta \eta} \right)
	-i\partial^{\rho}{\delta \over \delta{\cal A}^{\rho}}\Biggl]
	\itGamma[{\bm A};{\bm\eta},{\bm\tau}]-i\dbox{\cal B} }{
\hspace{30ex}
	-i(\delta^{\mu}_{\rho}\delta^{\nu}_{\sigma}
	+\delta^{\mu}_{\sigma}\delta^{\nu}_{\rho}- g^{\mu\nu} g_{\rho\sigma})
	\partial^{\rho} \! \! \left( \! \tau_{\mu\nu}^{\rm g}\partial^{\sigma}
	{\cal B}\! \right)=0\ . }
\ea	\lab{GWTA}
\ee
Owing to the above discussion, we need only the tree $\itGamma^{(0)}$ and the
one loop
$\itGamma^{(1)}$.

\paragraph{$\itTheta^{\mu\nu}_{\rm g}$-inserted part}:a typical
subdiagram is seen in  Fig.5.
$$ \epsfbox{Fig5.epsf} $$
{}From \eqn{GWTA},
\be
\Biggl[e \left( \overline{\eta} { \delta \over \delta\overline{\eta} }
	-\eta{\delta \over \delta \eta} \right)
	-i\partial^{\rho}{\delta \over \delta{\cal A}^{\rho}}\Biggl]
	\itGamma[{\bm A};{\bm\eta},{\bm\tau}]
	\Biggr\vert_{{\cal B}={\bm\tau}=0}=0\ .
\lab{GWTO}
\ee
Write the tree Green's functions of electron as
\be
\ba{l}
	\lllongeq{ \lefteqn{i(-e)^{n+1}I^{\rho\rho_{1}\cdots\rho_{n}}
	(q, q_{1},\cdots, q_{n}; k,p)} }
{ \equiv
	\int\!\!d^{4}yd^{4} z\prod^{n}_{j=1}d^{4}x_{j}
	\exp[iky-ipz+i\sum^{n}_{j=1}q_{j}x_{j}] }
{  \times{\delta^{2}\over \delta\overline\eta(y)\delta\eta(z)}
	{i\delta^{n+1}\itGamma^{(0)}[{\bm A};{\bm\eta},{\bm\tau}]
	\over\delta {\cal A}_{\rho}(0)\delta {\cal A}_{\rho_{1}}(x_{1})
	\cdots\delta {\cal A}_{\rho_{n}}(x_{n})}
	\Biggr\vert_{{\bm A}={\bm\eta}={\bm\tau}=0} }
{ \equiv  \sum_{\scriptstyle permutation\atop
	{\left\{{q\choose\rho},{q_{1}\choose\rho_{1}},
	\cdots,{q_{n}\choose\rho_{n}}\right\}}} \quad
\raisebox{-7ex}{\epsfbox{Fig6.epsf} }  \ , }
\ea \lab{AFL}
\ee
with $q=p-k-\sum q_{j}$. Therefore WT reads
\begin{equation}
	q_{\rho}I^{\rho\rho_{1}\cdots\rho_{n}}
	(q, q_{1},\cdots, q_{n}; k,p)
	=-I^{\rho_{1}\cdots\rho_{n}}(q_{1},\cdots,q_{n}; k+q,p)
	+I^{\rho_{1}\cdots\rho_{n}}(q_{1},\cdots,q_{n}; k,p-q)
	\lab{WTM}
\end{equation}
whose left hand side stands for the contribution from the gauge dependent part
in the photon propagator. The LSZ amplitude is obtained by sandwiching
\eqn{WTM} with
$\overline{u}(k,s)(\dslk -m)$ and $(\dslp -m)u(p,s')$. The gauge dependent
part is therefore becomes finally
\begin{equation}
	\overline u(k,s)(\dslk -m)q_{\rho}I^{\rho\rho_{1}\cdots\rho_{n}}
	(q, q_{1},\cdots, q_{n}; k,p) (\dslp -m)u(p,s')=0 \ ;
%	\lab{}
\end{equation}
since the right hand side of \eqn{WTM} cannot escape the cancellation because
of the
momentum shift $k+q$ or $p-q$.

For the one loop subgraphs define a Green's function,
\be
\ba{l}
\llongeq{
\lefteqn{(-e)^{n+1}\itPi^{\rho\rho_{1}\cdots\rho_{n}}(q,q_{1},\cdots,q_{n})}
}{ \equiv
	\int\!\!\prod^{n}_{j=1}d^{4}x_{j}
	\exp[i\sum^{n}_{j=1}q_{j}x_{j}]  {i\delta^{n+1}\itGamma^{(1)}[{\bm
A};{\bm\eta},{\bm\tau}]
	\over\delta {\cal A}_{\rho}(0)\delta {\cal A}_{\rho_{1}}(x_{1})
	\cdots\delta {\cal A}_{\rho_{n}}(x_{n})}
	\Biggr\vert_{{\bm A}={\bm\eta}={\bm\tau}=0} }
{ \equiv  \sum_{\scriptstyle permutation\atop
	{\left\{{q_{1}\choose\rho_{1}},
	\cdots,{q_{n}\choose\rho_{n}}\right\}}} \quad
\raisebox{-9ex}{\epsfbox{Fig7.epsf}}  \quad \ ,  }
\ea 	\lab{AFLO}
\ee
where $q=-\sum q_{j}$. WT reads
\begin{equation}
	q_{\rho}\itPi^{\rho\rho_{1}\cdots\rho_{n}}(q,q_{1},\cdots,q_{n})
	=0 \
	\lab{WTML}
\end{equation}
which tells us that gauge dependence also disappears.

\paragraph{$\itTheta^{\mu\nu}_{\rm m}$-inserted part}:
there are two type of typical subdiagrams with a $\itTheta^{\mu\nu}_{\rm m}$
insertion in Fig.8.
$$ \epsfbox{Fig8.epsf} $$
 In this case WT reads
\be
\Biggl[e \left( \overline{\eta} { \delta \over \delta\overline{\eta} }
	-\eta{\delta \over \delta \eta} \right)
	-i\partial^{\rho}{\delta \over \delta{\cal A}^{\rho}}\Biggl]
	{\delta \itGamma[{\bm A};{\bm\eta},{\bm\tau}]
	\over\delta\tau_{\mu\nu}^{\rm m}}=0 \ .
%	\lab{}
\ee
Define $\itTheta^{\mu\nu}_{\rm m}$
inserted Green's functions in the tree order:
\be
\ba{l}
\lllongeq{	\lefteqn{(-e)^{n+1}I^{\mu\nu;\rho\rho_{1}\cdots\rho_{n}}
	(q,q_{1},\cdots,q_{n}; k,p)} }
{ \equiv
	\int\!\!d^{4}yd^{4}zd^{4}x\prod^{n}_{j=1}d^{4}x_{j}
	\exp[iky-ipz+iqx+i\sum^{n}_{j=1}q_{j}x_{j}] }
{ \times{\delta^{2}\over \delta\overline\eta(y)\delta\eta(z)}
	{\delta\over\delta\tau_{\mu\nu}^{\rm m}(0)}
	{\delta^{n+1}\itGamma^{(0)}[{\bm A};{\bm\eta},{\bm\tau}]
	\over\delta {\cal A}_{\rho}(x)\delta {\cal A}_{\rho_{1}}(x_{1})
	\cdots\delta {\cal A}_{\rho_{n}}(x_{n})}
	\Biggr\vert_{{\bm A}={\bm\eta}={\bm\tau}=0}  }
{ \equiv  \sum_{\scriptstyle permutation\atop{\left\{{q\choose\rho},
	{q_{1}\choose\rho_{1}},
	\cdots,{q_{n}\choose\rho_{n}}\right\}}}
	\sum_{\scriptstyle\ all\ possible\atop\scriptstyle insertions
	\ of \  \itTheta^{\mu\nu}_{\rm m}} \quad \raisebox{-5ex}{\epsfbox{Fig9.epsf} }
\ ,  }
\ea         \lab{b66}
\ee
and in the one loop:
\be
\ba{l}
\llongeq{	\lefteqn{-i(-e)^{n+1}\itPi^{\mu\nu;\rho\rho_{1}\cdots\rho_{n}}
	(q,q_{1},\cdots,q_{n})}  }
{   \equiv
	\int\!\!d^{4}x\prod^{n}_{j=1}d^{4}x_{j}
	\exp[iqx+i\sum^{n}_{j=1}q_{j}x_{j}]
	{\delta\over\delta\tau_{\mu\nu}^{\rm m}(0)}
	{\delta^{n+1}\itGamma^{(1)}[{\bm A};{\bm\eta},{\bm\tau}]
	\over\delta {\cal A}_{\rho}(x)\delta {\cal A}_{\rho_{1}}(x_{1})
	\cdots\delta {\cal A}_{\rho_{n}}(x_{n})}
	\Biggr\vert_{{\bm A}={\bm\eta}={\bm\tau}=0}  }
	{  \equiv \sum_{\scriptstyle permutation\atop
	{\left\{{q_{1}\choose\rho_{1}},
	\cdots,{q_{n}\choose\rho_{n}}\right\}}}
	\sum_{\scriptstyle\ all\ possible\atop\scriptstyle insertions
	\ of \  \itTheta^{\mu\nu}_{\rm m}} \quad \raisebox{-8ex}{\epsfbox{Fig10.epsf}
}\ .  }
\ea %	\lab{}
\ee
WT for $I^{\mu\nu}$'s and $\itPi^{\mu\nu}$'s are given as the
same as
\eqn{WTM} and \eqn{WTML};
\be
\ba{l}
\longeq{q_{\rho}I^{\mu\nu;\rho\rho_{1}\cdots\rho_{n}}
	(q,q_{1},\cdots,q_{n}; k,p)}
	{=-I^{\mu\nu;\rho_{1}\cdots\rho_{n}}(q_{1},\cdots,q_{n};k+q,p)
	+I^{\mu\nu;\rho_{1}\cdots\rho_{n}}(q_{1},\cdots,q_{n}; k,p-q)\ ,}
\ea   \lab{WTMT}
\ee
\begin{equation}
	\hspace{-15em}{q_{\rho}\itPi^{\mu\nu;\rho\rho_{1}\cdots\rho_{n}}
	(q,q_{1},\cdots,q_{n})=0}\ .
	\lab{WTMLT}
\end{equation}
\eqn{WTMLT} is considered as a generalization of the previous relation
\eqn{trv}.
{}From \eqn{WTMT} and \eqn{WTMLT} gauge dependence is wiped out again by a
similar
manner as in the $\itTheta^{\mu\nu}_{\rm m}$-inserted case.

Needless to say, ${G^{\mu \nu ; \lambda}_{\rm g}(q; k, p)}^{(1)}$ belong to the
case
\eqn{AFL} and ${G^{\mu \nu ; \lambda}_{\rm m}(q; k, p)}^{(1)}$ to \eqn{b66}.We
have
now convinced ourselves that the energy-momentum tensor
\eqn{b24} is gauge independent under the LSZ-formalism, which, in other words,
guarantees the passage from \eqn{b11} to \eqn{IIaselec}. The asymptotic
electron field
is therefore considered as gauge invariant but the relationship to the
interpolating field
is still unclear.

%%%%%%%%%%%%%%%%%%%%%%%%%%%%%%%%%%%%%%%%%%%%%%%%%
%%%%%%%%%%%%%%%%%%%%%%%%%%%%%%%%%%%%%%%%%%%%%%%%%
\section{Gauge Invariant Approaches}

According to discussions in the foregoing sections, the LSZ asymptotic states
is
gauge invariant with the use of the physical state conditions \eqn{pscB} and
\eqn{IIfrelec}.
However, it is preferable to introduce gauge invariant operators for photon and
electron, which
would guarantee gauge independence more directly.

To investigate that let us recall the gauge
invariant quantities in (classical) electrodynamics: the minimal coupling term,
\be
\overline{\psi}(x) i\gamma^\mu \big( \partial_\mu - ieA_\mu(x) \big) \psi(x) \
, \lab{g}
\ee
and the field strength tensor $F_{\mu\nu}(x)$.

The gauge transformation is expressed as
\be
\ba{l}
\llongeq{ A_\mu(x) \mapsto A_\mu(x) + \partial_\mu \chi(x) \ ,}
{\psi(x) \mapsto  e^{ie\chi(x)} \psi(x)  \ ,}
{\overline{\psi}(x) \mapsto  \overline\psi (x) e^{-ie\chi(x)} \  .}
\ea \lab{h}
\ee
In terms of the components, the photon part of \eqn{h} reads as
\be
\ba{l}
\longeq{ A_0(x) \mapsto  A_0(x) + \dot \chi(x) \ , }
{ {\bm A}(x) \mapsto  {\bm A}(x) - {\bm \nabla} \chi(x) \ .}
\ea  \lab{i}
\ee
Now we decompose the vector potential ${\bm A}(x)$ into
\be
{\bm A}(x) = {\bm A}_{\rm T}(x) + {\bm A}_{\rm L}(x) \ ,  \lab{k}
\ee
where $ {\bm A}_{\rm T}(x)(  {\bm A}_{\rm L}(x)) $ denotes the transverse
(longitudinal) component with respect to the derivative ${\bm \nabla}$; thus
\be
\ba{l}
\longeq{ {\bm \nabla}\cdot {\bm A}_{\rm T}(x) =   0 \ , }
{{\bm \nabla} \times {\bm A}_{\rm L}(x) =   0 \ .}
\ea  \lab{l}
\ee
In view of \eqn{h}, we then obtain the transformation rule:
\be
\ba{l}
\longeq{ {\bm A}_{\rm T}(x) \mapsto   {\bm A}_{\rm T}(x)  \ , }
{ {\bm A}_{\rm L}(x) \mapsto  {\bm A}_{\rm L}(x) - {\bm \nabla} \chi(x) \ . }
\ea  \lab{m}
\ee
{}From this we recognize that {\em the transverse part,} ${\bf A}_{\rm
T}(x)$,{\em  is  gauge
invariant} \cite{KT}. The vector ${\bm \nabla}$ sets up a reference axis along
which gauge
invariant quantities can be constructed. In order to find other invariant
quantities, let us go
back to
\eqn{g}. First it should be noticed that
\be
\ba{l}
\longeq{ \psi_{\rm inv}^{\rm C}(x) \equiv  \exp \Big[ ie\int^{\bm x} d{\bm z}
\cdot
{\bm A}_{\rm L}(x^0, {\bm z}) \Big] \psi(x) \  , }
{ \overline{ \psi}_{\rm inv}^{\rm C}(x) \equiv     \overline{ \psi}(x) \exp
\Big[
-ie\int^{\bm x} d{\bm z} \cdot {\bm A}_{\rm L}(x^0, {\bm z}) \Big] \  , }
\ea  \lab{n}
\ee
are gauge invariant under \eqn{i} and \eqn{m}, path-independent owing to
\eqn{l}, (hence the beginning point of the integral can be arbitrary), and {\em
in fact local}. This is essentially the Dirac's physical electron
\cite{DT,LMc}; since
\be
 {\bm A}_{\rm L}(x) = {\bm \nabla} { {\bm \nabla} \cdot
{\bm A}
\over {\bm \nabla}^2 }(x) \ ,
\ee
where
\be
{f  \over {\bm \nabla}^2}(x) \equiv  - {1 \over 4\pi } \int_{-\infty}^\infty
d^3 {\bm y}
 \  {f(x_0, {\bm y}) \over | {\bm x} - {\bm y} | }  \ ,
\ee
so that \eqn{n} becomes
\be
\ba{l}
\longeq{ \psi_{\rm inv}^{\rm C}(x) =  \exp  \!  \left[ie  { {\bm \nabla} \cdot
{\bm A} \over {\bm \nabla}^2 }(x) \right] \psi(x) \ , }
{ \overline{ \psi}_{\rm inv}^{\rm C}(x) =   \overline{ \psi}(x) \exp  \! \left[
-ie { {\bm \nabla} \cdot {\bm A} \over {\bm \nabla}^2 }(x) \right]  \ , }
\ea  \lab{dirac}
\ee
which is the Dirac's electron.

The minimal coupling term \eqn{g} becomes
\be
\overline{ \psi}_{\rm inv}^{\rm C}(x) i\Bigg[ \gamma^0\bigg\{ \partial_0 -
ie\bigg(
A_0(x) +
\int^{\bm x} d{\bm z} \cdot \dot{ {\bm
A}_{\rm L} }(x^0, {\bm z}) \bigg) \bigg\} +
{\bm \gamma} \cdot \bigg( {\bm \nabla} + ie {\bm A}_{\rm T}(x) \bigg) \Bigg]
\psi_{\rm
inv}^{\rm C}(x) \ ,  \lab{o}
\ee
leading to the gauge invariant potential,
\be
A_\mu^{{}^{\rm C}}(x) \equiv \Big( A_0^{{}^{\rm C}}(x), - {\bm A}^{{}^{\rm
C}}(x) \Big)  \ ,  \lab{qa}
\ee
with ${\bm A}^{{}^{\rm C}}(x) \equiv {\bm A}_{\rm T}(x),$ and
\be
A_0^{{}^{\rm C}}(x) \equiv A_0(x) + \int^{\bm x} d{\bm z} \cdot \dot{ {\bm
A}_{\rm L} }(x^0, {\bm z})  \ .  \lab{p}
\ee
Apparently
\be
{\bm \nabla} \cdot {\bm A}^{{}^{\rm C}}(x)  = 0 \ .  \lab{r}
\ee

In view of \eqn{r} this is nothing but the Coulomb gauge. From this lesson, it
should be
recognized that {\em to set up gauge invariant operators is nothing but to fix
the gauge,}
whose result also can be seen in a covariant form by Steinmann \cite{ST}. We
should use the
terminology ``BRS invariance'' here instead of gauge invariance since we now
move into quantum
theory. The starting Lagrangian is the Nakanishi-Lautrup one \eqn{I8} and the
BRS
transformation is given by \eqn{a9}. (The Faddeev-Popov ghosts are irrelevant
all the time in
QED.) The BRS invariant fermion fields are thus defined by
\be
\ba{l}
	\longeq{ \itPsi(x) \equiv \psi_{\rm inv}^{\phi}(x) \equiv \exp
	\left[-ie\int\!\!d^{4}y\,\phi^{\mu}(x-y)A_{\mu}(y)\right]\psi(x)  \ , }
{ \overline\itPsi(x) \equiv \overline{\psi}_{\rm inv}^{\phi}(x)\equiv
	\overline\psi(x) \exp
	\left[ie\int\!\!d^{4}y\,\phi^{\mu}(x-y)A_{\mu}(y)\right] \ ,  }
\ea \lab{pf}
\ee
with a real distribution $\phi^{\mu}(x)$ satisfying
\begin{equation}
	\der{\mu}{}\phi^{\mu}(x)=\delta^{4}(x)\ .
	\lab{derphi}
\end{equation}
The minimal coupling term becomes
\be
	\overline\itPsi(x)\left[i\dslpart-m
	+e\gamma^{\nu}\!\int\!\!d^{4}y\,\phi^{\mu}(x-y)F_{\mu\nu}(y)\right]
	\itPsi(x)
\ee
so that the BRS invariant potential in the Steinmann's approach is
\be
\ba{l}
\llongeq{ A^{\phi}_\mu (x) \equiv - \int\!\!d^{4}y\,\phi^{\nu}(x-y)F_{\mu
\nu}(y) }{
=  A_{\mu}(x)-\partial_{\mu}^{x}\!\int\!\!d^{4}y\,\phi^{\nu}(x-y)
	A_{\nu}(y) }
{ = \int\!\!{d^{4}q\over(2\pi)^{4}}{\rm e}^{-iqx}
	(\delta^{\lambda}_{\mu}+iq_{\mu}\phi^{\lambda}(q))
	\int\!\!d^{4}y\,{\rm e}^{iqy}A_{\lambda}(y) \ ,  }
\ea \lab{c18}
\ee
where use has been made of the Fourier transformation
\begin{equation}
	\phi^{\mu}(x)=\int\!\!{d^{4}q\over(2\pi)^{4}}
	\phi^{\mu}(q){\rm e}^{-iqx}  \  .
\end{equation}
According to the second expression in \eqn{c18}, it can be regarded that
$A_\mu$ has been
decomposed into the gauge invariant part, $A^{\phi}_\mu$, and the variant part,
$\overline{A}_\mu(x)$;
\be
\ba{l}
\longeq{ A_{\mu}(x)= A^{\phi}_\mu (x) + \overline{A}_\mu(x) \ ; }
{\overline{A}_\mu(x) \equiv
\partial_{\mu}^{x}\!\int\!\!d^{4}y\,\phi^{\nu}(x-y){A}_\nu(y) \
, }
\ea
\ee
which corresponds to \eqn{k}. In this case, the reference vector is of course
$\phi^\mu$.

\eqn{derphi} implies
\begin{equation}
	q_{\mu}\phi^{\mu}(q)=i \ ,
	\lab{qphi}
\end{equation}
yielding to
\be
 q_{\mu} \Bigl( \delta^{\mu}_{\nu} + i q_{\nu}\phi^{\mu}(q) \Bigr) = 0 \ ;
\quad \phi^{\mu}(-q) = -\phi^{\mu}(q)  \ ,
\lab{c24}
\ee
which furthermore leads to a projection property:
\be
\Bigl( \delta^{\mu}_{\nu} + i q_{\nu}\phi^{\mu}(q) \Bigr) \Bigl(
\delta^{\nu}_{\lambda} + i
q_{\lambda}\phi^{\nu}(q) \Bigr)  = \Bigl( \delta^{\mu}_{\lambda} + i
q_{\lambda}\phi^{\mu}(q) \Bigr)   \ .
\ee

If
\begin{equation}
	\phi^{\mu}(x)
	= \Bigl( 0, { {\bm x} \over4\pi\vert{\bm x}\vert^{3}}
	\delta(x_{0}) \Bigr)  \ ,
	\lab{c19}
\end{equation}
\eqn{pf} becomes the Dirac's electron \eqn{dirac} and \eqn{c18} becomes
\eqn{qa}. Likewise,
if the support of $\phi^{\mu}$ lies in a space-like region BRS invariant
operators, \eqn{pf}
and \eqn{c18}, are well-defined by all means. On the contrary, if
$\phi^{\mu}$'s support
includes a time-like region, the expressions for $\itPsi$'s lose its meaning by
themselves
because of the non-commutativity of $A$'s  and $\psi$. However, even in this
case
perturbation can lay down the definition: for example, considering the
quantity,
\be
\bra{0}T^*{A^{\phi}}^{\lambda_{1}}(x_{1})\cdots {A^{\phi}}^{\lambda_{n}}(x_{n})
	\itPsi(y_{1})\cdots\itPsi(y_{m})
	\overline\itPsi(z_{1})\cdots\overline\itPsi(z_{m})
	\ket{0} , \lab{c20}
\ee
in terms of perturbation, imparts a meaning to those BRS
invariant operators.  \eqn{c20} can, however, be more simplified, in view of
\eqn{c18}, such
that
\be
\ba{l}
\llongeq{	\bra{0}T^*A^{\lambda_{1}}(x_{1})\cdots A^{\lambda_{n}}(x_{n})
	\itPsi(y_{1})\cdots\itPsi(y_{m})
	\overline\itPsi(z_{1})\cdots\overline\itPsi(z_{m})
	\ket{0} } {  =   \bra{0}
	T^*A^{\lambda_{1}}(x_{1}) \cdots A^{\lambda_{n}}(x_{n})
	\psi(y_{1})\cdots\psi(y_{m})
	\overline\psi(z_{1})\cdots\overline\psi(z_{m}) }
	{ \times \exp\left[-ie\!\int\!\!d^{4}s \,
	\phi^{\rho}(y_{1}-s)A_{\rho}(s)\right] \cdots \exp\left[ie\!\int\!\!d^{4}s\,
	\phi^{\rho}(z_{m}-s)A_{\rho}(s)\right]\ket{0}  \ . }
 \ea  \lab{gphi}
\ee
Therefore the effect of the physical electron is solely found in the additional
vertex, $\phi^{\mu}$-vertex as in  Fig.11.
$$ \epsfbox{Fig11.epsf} $$
Furthermore by noting that $\it\Psi$
simply turns out to be $\psi$ under the loops, tree graphs of
electron are only relevant. Especially for the two-point function,
\be
\ba{l}
\llllongeq{ \bra{0}T^*\itPsi(y)\overline\itPsi(z)\ket{0} }
{= \raisebox{-12.5ex}{ \epsfbox{Fig12.epsf} }}
{ = \int\!\!{d^{4}k\over(2\pi)^{4}}{\rm e}^{-ik(y-z)}
	\Biggl[{i\over\dslq-m} }
	{  +\int\!\!{d^{4}l\over(2\pi)^{4}}{i\over\dslq-m}ie\gamma_{\rho}
	{i\over\dslq-\dsll-m}ie\gamma_{\sigma}{i\over\dslq-m} }
 { \times 	{-i\over l^{2}}\Bigl\{g^{\rho\sigma}+i l^{\rho}\phi^{\sigma}(l)
	+ i\phi^{\rho}(l)l^{\sigma}
	- l^{\rho}l^{\sigma}\phi(l)\phi(l)\Bigr\}
	+O(e^{4})\Biggr]  \ . }
\ea
\ee
Note that the photon propagator has been replaced such that
\begin{equation}
	{-i\over l^{2}}d^{\rho\sigma}(l)\longrightarrow
	{-i\over l^{2}}\Bigl\{g^{\rho\sigma}+i l^{\rho}\phi^{\sigma}(l)
	+ i\phi^{\rho}(l)l^{\sigma}
	- l^{\rho}l^{\sigma}\phi(l)\phi(l)\Bigr\}  \ .
\lab{mp}
\end{equation}
This is our statement: the $\alpha$-dependent propagator turns into a
$\phi$-dependent one.

Similarly in a multi-point function
\eqn{gphi}, take a single fermion line to which n+1 vertices are attaching.
With regard to a
special vertex
$\rho$ out of which  the momentum $q$ flows, there arise two new contributions
from
$\phi^{\rho}(q)$. In terms of the notations in the previous section 2.4 it
gives
\be
\ba{l}
\lllllongeq{ i(-e)^{n+1}I^{\rho\rho_{1}\cdots\rho_{n}}
	(q,q_{1},\cdots,q_{n};k,p) }
{ + ie\phi^{\rho}(q) i(-e)^{n}I^{\rho_{1}\cdots\rho_{n}}
    (q_{1},\cdots,q_{n};k+q,p) }
{ +i(-e)^{n}I^{\rho_{1}\cdots\rho_{n}}(q_{1},\cdots,q_{n};k,p-q)
     ie\phi^{\rho}(q)  }
{ =  i(-e)^{n+1}\Bigl(\delta^{\rho}_{\lambda}+i\phi^{\rho}(q)q_{\lambda}\Bigr)
    I^{\lambda\rho_{1}\cdots\rho_{n}}(q,q_{1},\cdots,q_{n};k,p) }
{ = \sum_{\scriptstyle permutation\atop{\left\{{q\choose\rho},
	{q_{1}\choose\rho_{1}},
	\cdots,{q_{n}\choose\rho_{n}}\right\}}} \quad
\raisebox{-5ex}{\epsfbox{Fig13a.epsf}} }
{ + \sum_{\scriptstyle permutation\atop{\left\{{q_{1}\choose\rho_{1}},
	\cdots,{q_{n}\choose\rho_{n}}\right\}}}\left\{
\raisebox{-5ex}{\epsfbox{Fig13b.epsf} }
\right\}\ , }
\ea  \lab{Iphi}
\ee
where use has been made of WT about $I$'s \eqn{WTM} to the
second expression. By applying the same manipulation to each vertex, the
$n+1$-th
photon amputated part of \eqn{gphi} becomes to
\be
\ba{l}
\longeq{\!\!\! \eqn{gphi}  \longrightarrow
i(-e)^{n+1}\Bigl(\delta^{\rho}_{\lambda}+ i\phi^{\rho}(q)q_{{}_{\lambda}
}\Bigr)

\Bigl(\delta^{\rho_{1}}_{\lambda_{1}}+
i\phi^{\rho_{1}}(q_{1}){q_{1}}_{{}_{\lambda_{1}}
}\Bigr)
\!	\cdots \!
\Bigl(\delta^{\rho_{n}}_{\lambda_{n}}+
i\phi^{\rho_{n}}(q_{n}){q_{n}}_{{}_{\lambda_{n}}
}\Bigr) } { \hspace{53ex} \times
    I^{\lambda\lambda_{1}\cdots\lambda_{n}}
    (q,q_{1},\cdots,q_{n};k,p)  \ .  }
\ea
\ee
Accordingly we find that each photon index, $I^{\rho \cdots}$, is modified to
$\Bigl(\delta_\kappa^\rho  + i\phi^{\rho}(q)q_{{}_{\kappa} } \Bigr)I^{\kappa
\cdots}$ as
the result of adopting the gauge invariant electron $\it\Psi$, which, combined
with the
photon part \eqn{c18}, leads us to the result that the photon propagator is
modified to
\be
\ba{l}
	\longeq{ D_{ \mu \nu}^{\phi}(q) \equiv
\Bigl(\delta^{\rho}_{\mu}+i\phi^{\rho}(q)q_{\mu}\Bigr)
	{-i\over q^{2}}d_{\rho\sigma}(q)
	\Bigl(\delta^{\sigma}_{\nu}+ i\phi^{\sigma}(q)q_{\nu}\Bigr) }
{ \hspace{8ex} = {-i\over q^{2}} \Bigl\{g_{\mu\nu}+iq_{\mu}\phi_{\nu}(q)
	+ i\phi_{\mu}(q)q_{\nu}- q_{\mu}q_{\nu}\phi(q) \cdot \phi(q) \Bigr\}  \ ,  }
\ea \lab{mp2}
\ee
where  use has been made of the projection property \eqn{c24}. As expected,
{\em gauge
($\alpha$) dependence has been taken over by $\phi^\mu$.}

The covariant Landau gauge is realized
by choosing $\phi^{\mu}(l)$ as
\be
	\phi^{\mu}(l)={il^{\mu}\over l^{2}} \  ,
\ee
giving
\be
D_{ \mu \nu}^{L}(q) \equiv  {-i\over q^{2}} \left( g_{\mu\nu}- {q_{\mu}q_\nu
\over
q^2} \right)
\ .
\ee
Also the Coulomb gauge propagator is, in view of \eqn{c19}, given, by choosing
\begin{equation}
	\phi^{\mu}(l)= \Big( 0, {-i{\bm l} \over{\bm l}^{2} } \Big) \  ,
%	\lab{}
\end{equation}
as
\begin{equation}
	D_{ \mu \nu}^{C}(q) \equiv {-i\over l^{2}}\left(g_{ \mu \nu}
	+{ \sum_j \left( g_{\nu j} l_{\mu}l^{j}+g_{\mu j}l^{j}l_{\nu} \right)
	-l_{\mu}l_{\nu} \over {\bm l}^{2} } \right)  \  .
	\lab{c31}
\end{equation}

In this way, we have recognized that {\em building up gauge (BRS) invariant
electron
and photon is merely synonymous to fixing the gauge.}

Although all the operators in the expectation value,
\be
\bra{0}T \itTheta^{\mu\nu}(x) {A^{\phi}}^{\lambda_{1}}(x_{1})\cdots
{A^{\phi}}^{\lambda_{n}}(x_{n})
	\itPsi(y_{1})\cdots\itPsi(y_{m})
	\overline\itPsi(z_{1}) \cdots \overline\itPsi(z_{m})
	\ket{0}  \ ,
\ee
with $\itTheta^{\mu\nu}(x)$ being given in \eqn{b24}, are BRS invariant, the
value itself
depends thus on $\phi$. Even in this invariant approach there still need the
physical state
conditions \eqn{pscB} and \eqn{IIfrelec} for proving $\phi$-independence of
$\itTheta^{\mu\nu}(x)$.

In spite of the fact that the BRS invariant electron and photon are not so
useful
for the proof of gauge independence, they serve us as a probe into the
structure
of the theory. For example, the LSZ-mapping is easily clarified with the aid of
$\itPsi$:
\be
 \itPsi(x) \mapsto Z^{1/2}_2 \psi^{as}(x)  + \mbox{higher orders} \ ,
\lab{LSZe}
\ee
together with the photon sector
\be
\ba{l}
\longeq{ A_\mu\kk{x} \mapsto Z^{1/2}_3 A^{\,as}_\mu\kk{x} +  \mbox{higher
orders} \ , }
{ \ {B}\kk{x}\mapsto  Z_3^{-1/2} B^{\,as}\kk{x}    \ .}
\ea     \lab{LSZp}
\ee
Note that the relation \eqn{LSZe} could be established with the aid of the
invariant approach:
since from \eqn{IIaselec} the asymptotic electron has been confirmed as
BRS invariant.

As was stressed before, if the support of $\phi^\mu$ is space-like, like in the
case of the
Dirac's electron, \eqn{LSZe} can hold as the (weak) operator relation.
Therefore strictly
speaking, we can declare that the LSZ-mapping can be confirmed only in the case
of Dirac's
electron. This fact also tells us that in QED electron can behave as
observable, whose
statement could then be generalized to QCD as a trial for illustrating the
dynamical
mechanism of quark confinement \cite{LMc}.

%%%%%%%%%%%%%%%%%%%%%%%%%%%%%%%%%%%%%%%%%%%%%%%%%%%%%%%%
%%%%%%%%%%%%%%%%%%%%%%%%%%%%%%%%%%%%%%%%%%%%%%%%%%%%%%%%

\section{Physical States in Functional Representation}

In this section we discuss other physical states in terms of
the functional representation. Also by use of that we build up the path
integral formula in the Coulomb
gauge and make an explicit transformation to the covariant gauge. We clarify
the reason for this ability.

\subsection{Other Physical States}

Consider $A_0 =0$ gauge in the conventional treatment
\cite{JSCL}: all three components ${\bm A}$ are assumed dynamical and obey the
commutation relations,
\be
 [ \hat{A}_j({\bf x}), \hat{E}_k({\bf y}) ] = i\delta_{jk} \delta({\bf x} -
{\bf y}) ,
\quad    [\hat{A}_j({\bf x}), \hat{A}_k({\bf x})] = 0 = [\hat{E}_j({\bf x}),
\hat{E}_k({\bf x})] ; \quad (j, k = 1, 2, 3) \  .
\ee
Here and hereafter the caret designates operators. The physical state condition
\eqn{Igauss} is then
\be
\hat{\Phi}({\bm x})| phys \rangle \equiv
\bigg[ \sum_{k=1}^3 \big(
\partial_k \hat{{\bm E}}_k({\bm x}) \big) + J_0({\bm x}) \bigg] |phys \rangle =
0 \ ,   \lab{d2}
\ee
where $J_\mu(x)$ is supposed as a $c$-number current.  First this should be
read such
that {\em there is no gauge transformation in the physical space}. As was
mentioned in the introduction,
the representation of the physical state cannot be obtained within the usual
Fock space since
$\hat{\Phi}({\bm x})$ is a local operator to result in $\hat{\Phi}({\bm x})=0$
\cite{PR}, but can be in the
functional (Schr\"odinger) representation \cite{FJ}:
\be
\ba{c}
\longeq{ \widehat{\bm A}({\bm x}) | \{{\bm A} \} \rangle  =   {\bm A} ({\bm x})
|
\{{\bm A} \} \rangle  \  , \quad
\widehat{\bm E } ({\bm x})  | \{{\bm E} \} \rangle =  {\bm E} ({\bm x}) |
\{{\bm E} \}
\rangle \ ,  }{ \langle \{ {\bm A} \} | \widehat{ \bm E} ({\bm x}) =   -i {
\delta
\over {\delta  {\bm A}({\bm x}) } }\langle \{ {\bm A} \} |  \  ,  \dots }
\ea \lab{schrep}
\ee
To see the reason take the states, $| \{{\bm A} \} \rangle, | \{{\bm E} \}
\rangle$, which can
be constructed in terms of the Fock states. The creation and annihilation
operators are given by
\be
\ba{l}
\longeq{\widehat{\bm A}({\bm x})  =   \int {d^3 {\bm k} \over {(2 \pi)^{3/2}
\sqrt {2|
{\bm k}| } }} \big( {\bf a}( {\bm k} ) e^{i{\bs k} \cdot {\bs x} }   +   {\bf
a}^{\dag} (
{\bm k} ) e^{- i{\bs k} \cdot {\bs x} } \big) }{ [a_i({\bm k} ),
a_j^{\dag}({\bm k}')] =
\delta_{ij}\delta({\bm k}-{\bm k}'),
\quad  [a_i({\bm k} ), a_j({\bm k}')] = 0 \ , }
\ea \lab{expan}
\ee
with the vacuum $|0 \rangle$;
\be
 {\bf a}( {\bf k} ) |0 \rangle = 0 \ . \lab{vacuum}
\ee
Now recall the quantum mechanical case \cite{KAS}:
\be
\ba{c}
\longeq{ \hat q | q \rangle = q | q \rangle \ ,  \qquad  \quad \hat p | p
\rangle = p | p
\rangle \ , }{ \hat q = { 1 \over \sqrt2}  \left( a  + a^{\dag} \right) \ ,
\quad  \hat p = { 1
\over {\sqrt2}i} \left( a -  a^{\dag} \right) \  ;  \qquad  a |0\rangle =0 \  ,
}
\ea \lab{qprep}
\ee
then
\be
\ba{l}
\longeq{ | q \rangle  = {1 \over \pi^{1/4} }
 \exp \left( - { q^2 \over 2} + \sqrt2 q a^{\dag} - {( a^{\dag})^2 \over 2}
\right)  | 0
\rangle \  , }{ | p \rangle  = {1 \over \pi^{1/4} } \exp \left( - { p^2 \over
2} + \sqrt 2 ip
a^{\dag} + {  ( a^{\dag} )^2 \over 2 } \right)  | 0 \rangle \  .  }
\ea \lab{qprepa}
\ee
These bring  us to
\be
\ba{l}
\longeq{   | \{{\bm A} \} \rangle   \
\underline{\sim } \  \exp \Big[ -{1 \over 2} \int d^3 {\bm x} \ d^3{\bm y} \
{\bm A}(
{\bm x} ) K( {\bm x} - {\bm y} ) {\bm A}({\bm y} )}{ \hspace{7ex} + \int d^3
{\bm x}
\! \!  \int d^3 {\bm k}  \
\sqrt{ 2|{\bm k} |  \over  (2\pi )^3  }  \ {\bm A} ( {\bm x} ) \! \cdot \! {\bf
a}^{\dag} ({\bm k}
) e^{-i{\bs k} \cdot {\bs x} }
 - {1 \over 2} \int d^3 {\bm k}  \  {\bf a}^{\dag} ({\bm k} ) \! \cdot \!
{\bf a}^{\dag} (-
{\bm k} ) \Big] | 0 \rangle  \  , }
\ea \lab{arep}
\ee
where
\be
K({\bm x}) \equiv \int  { d^3 {\bm k} \over  (2\pi )^3 }  | {\bm k} | e^{i {\bs
k} \cdot {\bs x} }
 \ , \lab{kfun}
\ee
which is apparently divergent so we must introduce some cut-off. The physical
state in the functional
representation is thus found as
\be
\langle \{{\bm A} \} | \hat{\Phi}({\bm x})  |  phys \rangle  = \left(
-i {\bm \nabla}  {\delta \over  {\delta  {\bm A}({\bm x}) } }  - J_0({\bm x})
\right)
\Psi_{\rm phys}[A] = 0 \ ,  \lab{phystasol}
\ee
where
\be
\Psi_{\rm phys}[A]  \equiv \langle \{{\bm A} \} |   phys \rangle \ .
\ee

Now we can see the reason for having the physical state in this case. Within a
single Fock state the
physical state condition \eqn{d2} merely implies $\hat{\Phi}({\bm x})=0$ but
{\em the functional
representation consists of infinitely many collections of inequivalent Fock
spaces}:  since the inner
product of $| \{{\bm A} \}
\rangle$
\eqn{arep}, to the Fock vacuum is found to be
\be
\ba{l}
\longeq{ \langle \{{\bm A} \}  | 0 \rangle  \sim   \exp\left[  - { 1 \over 2}
\int d^3
{\bm x} \ d^3{\bm y} \ {\bm A}( {\bm x} ) \ K( {\bm x} - {\bm y} ) {\bm
A}({\bm y} )
\right] }
{ \hspace{8.5ex} = \exp\left[  - { 1 \over 2} \int { d^3 {\bm k} \over  (2\pi
)^3 } \
| {\bm k} | \ {\bm A}( {\bm k} ) {\bm A}(-{\bm k} ) \right]
\longrightarrow  0
 \ ; {}^{\forall}\!\{{\bm A} \}\ ,  }
\ea    \lab{inner}
\ee
when the cut-off becomes infinity. This happens in {\em any value of} ${\bm
A}({\bm x})$. Therefore (apart from the mathematical rigorousness of that) any
local first class constraint can be realized by means of the functional
representation. Furthermore the fact that the functional representation
contains an
infinite set of the Fock states enables us to perform an explicit gauge
transformation and prove gauge independence without recourse to any physical
state conditions.

\subsection{Proof of Gauge Independence by Path Integral}

Recall that the path integral formula can be
obtained with the aid of the functional representation. It then might be easily
convinced that {\em we can move freely from one gauge to another
in the path integral} \cite{AL,KS}.

Take the Coulomb case. The Hamiltonian
is given by
\be
H = \int d^3{\bm x} \left [ {1 \over 2} {{\bm E}_{\rm T}}^2 + {1 \over 2} ({\bm
\nabla}_i {\bm A}_{\rm T})^2 + {1 \over 2} J_0 {1 \over {\bm \nabla}^2 } J_0
-  {\bm
J} \cdot {\bm A}
\right] \  ,
\lab{colhamil}
\ee
where the third term in the right hand side is the nonlocal Coulomb energy term
with $J_\mu $ being assumed as $c$-number sources. The equal
time commutation relations are
\be
[ \hat{A}_i(x), \hat{E}_j(y)] = i\left( \delta_{ij} - {{\bm \nabla}_i {\bm
\nabla}_j \over
{\bm
\nabla}^2} \right) \! \! ({\bm x}, {\bm y}) \equiv i {\bf P}_{ij} ({\bm x},
{\bm y}) \ ,
\quad [\hat{A}_i, \hat{A}_j] = [\hat{E}_i, \hat{E}_j] = 0 \ ,
\lab{commt}
\ee
where
\be
{\bf P}_{ij} ({\bf x}, {\bf y}) = \int d^3 {\bm k} \  e^{i {\bs k} \cdot ({\bs
x}-{\bs y}) }
\left(
\delta_{ij} - {k_ik_j \over {\bm k}^2 } \right) \ ,  \lab{commta}
\ee
which can be diagonalized \cite{Tab} by means of ${\bm S}$ such that
\be
{\bm S} {\bm P} {\bm S}^T = \left({\matrix{1&&\cr &1&\cr &&0\cr}}\right),
\ee
where $T$ designates the transpose. Explicitly
\be
{\bm S}_{1i} = {\bm n}_i \ , \quad {\bm S}_{2i} = \left( {{\bm \nabla} \over
|{\bm \nabla}| } \times
{\bm n} \right)_{\!\! \! i}  \ , \quad {\bm S}_{3i} = {{\bm \nabla}_i \over
|{\bm \nabla}| } \ ,
\lab{elemes}
\ee
where ${\bm n}$ is some vector perpendicular to ${\bm \nabla}$; ${\bm
\nabla}\cdot {\bm n} =0$.
Owing to ${\bm S}$ two components can be picked out so that we can write
(omitting
the caret)
\be
 ({\bm S} {\bm A} )_a =
\widetilde{A}_a \ ,  \quad  ({\bm S} {\bm E} )_a = \widetilde{E}_a \  , \qquad
( a= 1,2) \  ,
\lab{cca}
\ee
 and
\be
 [ \widetilde{A}_a, \widetilde{E}_b ] = i\delta_{ab} \ ,  \quad (a, b = 1,2) \
{}.
\lab{ccb}
\ee
Also by noting that
\be
{\bm E}_{\rm T}^2 = {\bm E}_i{\bm P}_{ij} {\bm E}_j =  (\widetilde{E}_a)^2 \ ,
\ee
we obtain
\be
H(t) = \int d^3{\bm x} \left [ {1 \over 2} (\widetilde{E}_a)^2 + {1 \over 2}
({\bm \nabla}_i \widetilde{A}_a )^2 + {1 \over 2} J_0 {1 \over {\bm \nabla}^2 }
J_0   -
\widetilde{J}_a \widetilde{A}_a \right] \ , \lab{ccc}
\ee
where $\widetilde{J}_a \equiv ({\bm S} {\bm J} )_a$ and to specify the explicit
time
dependence through the Coulomb energy term we have written the Hamiltonian as
$H(t)$. The summation convention for the repeated indices must be implied.

The starting point of the path integral is \cite{KS},
\be
Z (T)    \equiv \lim_{N \rightarrow \infty} \left( {\bf I} -
i \Delta t H_N \right)\left( {\bf I} - i \Delta t H_{N-1} \right) \cdots \left(
{\bf I} - i
\Delta t H_1 \right)  \ ,     \lab{ccd}
\ee
where $ \Delta t \equiv T/ N$ and $H_j \equiv H(j\Delta t)$. (Usually the
Euclidean technique, $T \rightarrow -iT$, must be used \cite{KS}. Here in order
to
illustrate the way to get the path integral expression we keep $i$ in the trace
formula. Also to make a whole discussion well-defined it is necessary to
discretize
space ${\bm x}$ but in the following the continuum expression is employed only
for the notational simplicity.) The essential ingredients\footnote{Path
integral
expressions for relativistic field for instance, $\phi(x)$, obtained via
holomorphic
(canonical coherent) representation \cite{FS} in terms of the creation and the
annihilation operator $a({\bm k})^\dagger , a({\bm k})$, must suffer from
nonlocality whenever going back to the $\phi(x)$-representation, due to the
mixing of particle and anti-particle. In order to get rid of this difficulty,
we
should start with the field $\phi(x)$-diagonal representation which is nothing
but the functional Schr\"odinger one.} are the functional  (Schr\"odinger)
representation
\eqn{schrep} together with
\be
\ba{l}
\longeq{\int {\cal D} \widetilde{A}_a({\bm x})   | \{\widetilde{A}\} \rangle
\langle
\{\widetilde{A}\} | = {\bf I} \ , }{ \int {\cal D} \widetilde{E}_a({\bm x})  |
\{\widetilde{E}\}
\rangle  \langle  \{\widetilde{E}\} | = {\bf I} \ ; }
\ea  \lab{cce}
\ee
\be
\langle
\{\widetilde{E}\} | \{\widetilde{A}\} \rangle = \left(\prod_{{\bm x} } {1 \over
2\pi}
\right) \exp
\left[- i\int d^3 {\bm x}
\widetilde{E}_a({\bm x})\widetilde{A}_a({\bm x}) \right] \ ,  \lab{ccf}
\ee
whose (infinite) constant will be absorbed in the functional measure in the
following. Inserting \eqn{cce} into \eqn{ccd} successively and using
\eqn{ccf}, we obtain the path integral expression,
\be
\ba{l}
\llllongeq{ Z(T) =  \lim_{N \rightarrow \infty} \prod_{j=1}^N \int {\cal D}
\widetilde{A}_a({\bm x}; j) {\cal D}
\widetilde{E}_a({\bm x}; j) \exp \Biggl[ i \Delta t \int  d^3 {\bm x}  }
{ \hspace{6ex} \times  \Biggl\{   \widetilde{E}_a({\bm x}; j) { \left(
\widetilde{A}_a({\bm x}; j) -
\widetilde{A}_a({\bm x}; j-1) \right)  \over \Delta t }
  -   {1 \over 2} \left( \widetilde{E}_a({\bm x}; j) \right)^2  \!\!  + {1
\over 2} \left({\bm
\nabla}_i
\widetilde{A}_a({\bm x}; j) \right)^2 }
{  \hspace{9ex}  +  {1 \over 2} \int d^3 {\bm y} \ J_0({\bm x}; j) {1
\over {\bm \nabla}^2 }({\bm x}, {\bm y} ) J_0 ({\bm y}; j)  -
\widetilde{J}_a({\bm x}; j)
\widetilde{A}_a({\bm x}; j)  \Biggr\}  \Biggr]   }
{ \hspace{6ex} =   \int {\cal D} \widetilde{A}_a(x) {\cal D} \widetilde{E}_a(x)
\exp \Biggl[ i\int  d^4 x \biggl\{ \widetilde{E}_a\dot{\widetilde{A}}_a  }
{ \hspace{15ex}   - \left( {1 \over 2} \left(\widetilde{E}_a\right)^2  + {1
\over 2}
\left({\bm \nabla}_i
\widetilde{A}_a\right)^2 + {1 \over 2} J_0 {1 \over {\bm \nabla}^2 }J_0   -
\widetilde{J}_a
          \widetilde{A}_a \right) \biggr\}  \Biggr]  \ , }
\ea   \lab{ccg}
\ee
where we have again employed a continuous expression in the final line. (The
periodic boundary condition $\widetilde{A}_a({\bm x}; T) = \widetilde{A}_a({\bm
x}; 0)$ is now irrelevant so we have not specified it.) Now inserting
\be
1 =  \int {\cal D}\widetilde{A}_3 {\cal D}
\widetilde{E}_3 \delta( \widetilde{A}_3) \delta( \widetilde{E}_3)
\ee
into \eqn{ccg} and changing the variables to the original ${\bm A}$ and ${\bm
E}$
in view of \eqn{elemes} and \eqn{cca}, we obtain
\be
\ba{l}
\longeq{
 Z(T)  = \int {\cal D} {\bm E} \ {\cal D}{\bm A} \  ( \det
{\bm \nabla}^2) \delta({\bm \nabla}\! \cdot \! {\bm E})
\delta({\bm \nabla} \! \cdot \!  {\bm A}) \exp \! \! \Bigg[ i\int  d^4 x \Big\{
{\bm
E}\cdot
\dot{\bm A}   }{ \hspace{20ex} - \big( {1 \over 2} {\bm E}^2  + {1 \over 4}
(F_{ij})^2  +
{1
\over 2} J_0 {1 \over {\bm \nabla}^2 }J_0   -  {\bm J} \cdot {\bm A}\big)
\Big\}  \Bigg] \ .}
\ea   \lab{cch}
\ee
With the use of Fourier transformation of the delta function,
\be
\delta({\bm \nabla}\! \cdot \! {\bm E}) = \int {\cal D} \beta \exp \left[ i\int
d^4x \
\beta \ {\bm \nabla} \! \cdot \!  {\bm E} \right]  \ ,
\ee
the integration with respect
to ${\bm E}$ can be performed to obtain
\be
\ba{l}
\longeq{
 Z(T)  = \int {\cal D}{\bm A} \ {\cal D}\beta \  ( \det {\bm \nabla}^2)
\delta({\bm \nabla} \! \cdot \! {\bm A})  }{ \hspace{8ex}  \times \exp \Bigg[
i\int  d^4
x
\Big\{
  {1 \over 2} ( {\dot{\bm A}}  - {\bm \nabla} \beta )^2   -  {1 \over 4}
(F_{ij})^2  - {1
\over 2} J_0 {1
\over {\bm \nabla}^2 }J_0   +  {\bm J} \cdot {\bm A}  \Big\}  \Bigg] \ . }
\ea \lab{cci}
\ee
Here by reviving $A_0$ in the form of
\be
A_0 =  \beta  + {J_0 \over {\bm \nabla}^2}
\ee
the nonlocal (as well as instantaneous) Coulomb interaction is eliminated to
leave the
final form;
\be
 Z(T) = \int {\cal D}A_\mu \  ( \det {\bm \nabla}^2)
\ \delta({\bm \nabla} \! \cdot \!  {\bm A})  \exp \left[- i\int  d^4 x \Big\{
     {1 \over 4} F_{\mu\nu} F^{\mu \nu}  + J^\mu A_\mu  \Big\}  \right]  \ .
\lab{ccj}
\ee
Here using the relation,
\be
\delta({\bm \nabla} \! \cdot \!  {\bm A})  \sim  \lim_{\alpha \rightarrow 0}
\exp
\left[ -{i \over 2
\alpha } \int d^4 x \  ({\bm \nabla} \! \cdot \!  {\bm A} )^2 \right] ,
\ee
to \eqn{ccj} then integrating with respect to the gauge fields we obtain
\be
 Z(T) =  \exp \left[  i \int d^4 x \ d^4 y \ {1 \over 2} J^\mu(x)  D_{\mu
\nu}(x-y)
J^\nu(y)
\right]  ,  \lab{partition}
\ee
where $D_{\mu \nu}(x)$ is the Fourier transformation of the propagator
\eqn{c31}.

It is now a simple task to go to another gauge \cite{KS}. Suppose the new gauge
condition is
given by
\be
\partial^\mu {A}_\mu'(x) = f(x)  \  ,   \lab{cck}
\ee
where $f(x)$ is an arbitrary function. The gauge transformation is
\be
{A}_\mu' (x) = A_\mu(x) + \partial_\mu \chi(x)  \  .  \lab{ccl}
\ee
In order to find such $\chi(x)$, first we rewrite \eqn{cck} as
\be
 \dot{A}_0'(x)  = f(x) - {\bm \nabla}\cdot {\bm A'} (x)  \  ,
\ee
and substituting \eqn{ccl} into this to find
\be
\dbox \chi(x) = f(x) - \dot{A }_0(x)   \   .   \lab{ccm}
\ee
Thus the Jacobian is read as
\be
\det\left( {\delta A'_\mu \over \delta A_\nu} \right) = \det \left(
\delta_\mu^\nu - \delta_0^\nu {\partial_\mu \partial_0  \over \dbox } \right)
= \det
( -
{\bm \nabla}^2
\dbox^{-1} )   \   ,   \lab{ccn}
\ee
giving
\be
{\cal D}A_\mu  ( \det {\bm {\bm \nabla}}^2)  \delta({\bm \nabla} \! \cdot \!
{\bm
A})  = {\cal D}{A'}_\mu (\det
\dbox)\delta (\partial^\mu A'_\mu - f )   \   ,
\ee
where the minus sign in the determinant is irrelevant so we have dropped it.
Therefore
\be
Z(T) = \int {\cal D}A_\mu  ( \det \dbox)
\delta(\partial^\mu A_\mu - f )  \exp \Bigg[- i\int  d^4 x \Big\{
     {1 \over 4} F_{\mu\nu} F^{\mu \nu}  + J^\mu A_\mu  \Big\}  \Bigg]   \  .
\lab{cco}
\ee

In this way, the gauge transformation in the path integral expression can be
performed straightforwardly according to the fact that {\em the functional
representation has infinitely many
inequivalent representations}.

\section{Discussion}

In this paper, we recognize that the Belinfante's energy-momentum
tensor is gauge independent for all orders under the LSZ asymptotic
conditions in \S2.  Meanwhile, we know that to pick up the gauge invariant
electron or photon is merely synonymous to fix the gauge: the transverse part,
${\bm A}_{\rm T}, {\bm
\nabla}\cdot{\bm A}_{\rm T} = 0$, is gauge invariant; which is equivalent to
the
Coulomb gauge. In this case, $\psi$ itself becomes gauge invariant. It has been
sometimes argued that gauge symmetry is not a symmetry rather a
redundancy \cite{NP}. There need only two components but in order to recover
the rotational as well as the Lorentz invariance, spurious two components have
been added. These spurious components move around under
gauge transformations leaving the physical component unchanged. Therefore,
our observation in \S3 is natural; that is, picking up the gauge invariant
quantities leaves the Lorentz or rotational non-invariance. (If the Lorentz
invariance is kept respectable, the negative metric must be introduced, so that
operators themselves lose their significance without recourse to physical state
conditions\cite{JRa}.)  The situation reminds us of that of lattice gauge
theory
\cite{WC}; in which the gauge invariance has been maintained at the sacrifice
of the Lorentz (Euclidean) as well as the rotational invariance.  The method
provides us the non-perturbative treatment in a gauge invariant way, leading to
confinement in terms of an area law
\cite{WC} by the help of an  analogy with the critical phenomena in statistical
mechanics.  However, more physical and concrete views must be necessary in
order to understand the confinement problem thoroughly: for example, the
existence of Dirac's physical electron assures us that there is no confinement
in
QED.  Therefore if proof could be given in QCD that the physical quark fields
cannot be built up, the issue is resolved. The Gribov ambiguity \cite{GA} would
be the cornerstone of the proof: canonical commutation relations
(as a result of gauge fixing) between gauge fields can only hold within some
small region around a point, where the coupling constant is small enough.
Enlarging the region, we see that there takes part another gauge configuration,
bringing us to the impossibility of observing quarks \cite{LMc}.

According to the discussion in \S4 the path integral formulation would be
most suitable for treating gauge transformation; since infinitely many
Hilbert spaces compose the functional representation which is the basic of
the path integral formula. The issue is then how to patch those small regions
together to cover the whole functional space. There might be some hints from
the
recent observations in quantum mechanics on nontrivial manifolds \cite{TO} and
the path integral formula for a generic constraint \cite{ka}.

\vspace{1cm}
\begin{center}
{\large \bf Acknowledgments}
\end{center}

T. K. has got profound questions on gauge invariance from Y. Takahashi,
which was the starting point of this work and is grateful to N. Nakanishi
for his guidance to the covariant canonical LSZ formalism. Discussions with D.
McMullan are beneficial. The authors also thank to H. Yonezawa for his
calculation of the energy-momentum tensor on the violation of the
Ward-Takahashi relation.

\appendix

\section{The Covariant LSZ-Formalism in QED}
In this appendix we review the asymptotic behavior of photon fields
and the LSZ reduction formula given by Nakanishi \cite{NN} for a self-contained
purpose.

\subsection{Asymptotic Photon Field}
In order to know the behavior of asymptotic fields, it is necessary
to investigate the Heisenberg fields. Then start with
the Nakanishi-Lautrup Lagrangian \eqn{I8} in a renormalized form,
\begin{equation}
{\cal L}= -{1\over4}Z_{3}{F}^{\mu\nu}{F}_{\mu\nu} + \bpsi
({i\over2}\lrderpartb +eZ_{3}^{1\over2}\dslA -m) \psi
- A^{\mu}\der{\mu}{}B + {\alpha\over2}B^2 \ ,   \lab{Alag}
\end{equation}
where all quantities have been assumed renormalized except $\psi$, $m$, and
$e$, and
$Z_{3}$ is the wave function renormalization constant for photon.  The
equations
of motion  are
\be
\ba{l}
 \llongeq{ \dbox{A}_\mu-\KK{1-{\alpha}}\der{\mu}{}{B}={\bm j}_{\mu}\ , }
    { \der{}{\mu}{A}_\mu + {{\alpha B}}=0 \ , }
{ \dbox{B}=0 \ , }
 \ea \lab{EMB}
\ee
where
\begin{equation}
      {\bm j}_{\mu}= {Z_3}^{-{1\over2}}j_{\mu}
      -\KK{1-{Z_3}^{-1}}\der{\mu}{}{B}\ ,
%  	\lab{}
\end{equation}
and
\begin{equation}
	j_{\mu} \equiv  -e\bpsi\gamma_{\mu}\psi\ .
%	\lab{}
\end{equation}
The four-dimensional commutation relations among
$A_{\mu}$'s and $B$ are found as
\be
\ba{l}
  \llongeq{    \commu{{A}_\mu\kk{x}}{{B}\kk{y}} =
      -i\der{\mu}{x}D\kk{x-y} \ , }
  { \commu{{B}\kk{x}}{{B}\kk{y}} =0 \  ,}
{ \commu{B\kk{x}}{j_{\mu}\kk{y}} = 0 \ , }
  \lab{Bj}
\ea
\ee
where $D(x)$ is the invariant delta function,
\begin{equation}
   D\kk{x} \equiv \inte{\frac{d^4p}{\kk{2\pi}^3i}}
   \epsilon\kk{p_0}
   \delta\kk{p^2}\,\rmchar{e}^{-ipx}\ .
   \lab{DD}
\end{equation}
In order to obtain those for $A_{\mu}$'s compute first
$\bra{0}j_{\mu}\kk{x}j_{\nu}\kk{y}\ket{0}$ and then
utilize \eqn{EMB}, \eqn{Bj}, as well as
\begin{equation}
	\commu{{A}_k\kk{x_0\,,{\bm x}}}{{\mathop{A_{l}}^{\displaystyle .}}
	\kk{x_0\,,{\bm y}}}
	      =-\frac{i}{Z_3}g_{kl}\delta^3\kk{{\bm x}-{\bm y}} \  ,
%	\lab{}
\end{equation}
to find \cite{NoNa}
\begin{eqnarray}
  \lefteqn{\bra{0}\commu{{A}_\mu\kk{x}}{{A}_\nu\kk{y}}\ket{0}}
  \nonumber\\
    &\!=\!&\!-i\KK{g_{\mu\nu}-K\der{\mu}{}\der{\nu}{}}D\kk{x-y}
       +i\KK{1-{\alpha}}\der{\mu}{}\der{\nu}{}E\kk{x-y}
       \nonumber\\
    & &\mbox{\quad}-\frac{i}{Z_3}\integ{+0}{\infty}{ds}\,
       \sigma\kk{s}\KK{g_{\mu\nu}
       +\frac{\der{\mu}{}\der{\nu}{}}{s}}\Delta\kk{x-y;s}\ ,
    \lab{AA}
\end{eqnarray}
where
\begin{equation}
	K \equiv \frac{1}{Z_3}\integ{+0}{\infty}{ds}\frac{\sigma\kk{s}}{s}\ ,
%	\lab{}
\end{equation}
\begin{equation}
    Z_3 \equiv 1-\integ{+0}{\infty}{ds}\sigma\kk{s}\ ,
\end{equation}
with $\sigma\kk{s}$ being the spectral function, and $\Delta\kk{x;s}$ and
$E\kk{x}$ are expressed as
\be
\ba{l}
\longeq{ \Delta\kk{x;s}\equiv \inte{\frac{d^4p}{\kk{2\pi}^3i}}\epsilon\kk{p_0}
\delta\kk{p^2-s}\,\rmchar{e}^{-ipx}\ ,
}
     { E\kk{x} \equiv \inte{\frac{d^4p}{\kk{2\pi}^3i}}\epsilon\kk{p_0}
      \delta^\prime\kk{p^2}\,\rmchar{e}^{-ipx}\ ;  \quad
\delta^\prime\kk{a}\equiv {d \over
da}\delta\kk{a}  } \ .
\ea      \lab{E}
\ee

Once the four-dimensional commutation relations is obtained, so can be those
for the
asymptotic fields, ${A_\mu^{as}}\kk{x}, \ B^{\,as}\kk{x}$,
by simply throwing away the continuous spectrum part in \eqn{Bj} and \eqn{AA}:
\be
\ba{l}
  \llongeq{  \commu{{A_\mu^{as}}\kk{x}}{{A_\nu^{as}}\kk{y}}
    = -i\KK{g_{\mu\nu}-K\der{\mu}{}\der{\nu}{}}D\kk{x-y}
                       +i\KK{1-{\alpha}}\der{\mu}{}\der{\nu}{}E\kk{x-y}\ ,}
    { \commu{{A_\mu^{as}}\kk{x}}{B^{\,as}\kk{y}} =
    -i\der{\mu}{x}D\kk{x-y}\ ,
   }
   {  \commu{B^{\,as}\kk{x}}{B^{\,as}\kk{y}} = 0 \  .
    }
\ea  \lab{Aasycomm}
\ee
{}From this the equations of motion for the asymptotic fields reads
\be
\ba{r}
\llongeq{  \dbox{A_\mu^{as}}-\KK{1-{\alpha}}\der{\mu}{}
     B^{\,as}=0 \ , }
{ \der{}{\mu}{A_\mu^{as}} + {\alpha} B^{\,as}=0 \ , }
{  \dbox B^{\,as}=0 \ .  }
 \ea    \lab{AEq}
\ee

It should be noted that the canonical structure \eqn{Aasycomm} differs from
that
of the free theory in which the equations of motion is the same as \eqn{AEq}
but
the commutation relations is given without $K$! The Lagrangian leading to
\eqn{Aasycomm} as well as \eqn{AEq} is found as
\be
{\cal L}^{as}= -{1\over4} {F^{as}}^{\mu\nu} {F^{as}}_{\mu\nu}  -
{A^{as}}^{\mu}\der{\mu}{}B^{as} +
 {K \over 2} \partial_\mu B^{as} \partial^\mu B^{as} +
{\alpha\over2}\left(B^{as}\right)^2 \ .
\ee
Note the existence of the kinetic term of $B$.

\subsection{Wave Functions}

To obtain the LSZ formula, there needs to construct wave functions. When
$\alpha\ne1$ three
sets of positive frequency functions
$\KKK{{h_{{\bs k}\sigma}}^{\!\!\mu}\kk{x}}$,
$\KKK{{f_{{\bs k}\sigma}}^{\!\!\mu}\kk{x}}$ and
$\KKK{{g_{\bs k}}\kk{x}}$ must be prepared to meet the equations,
\be
\ba{l}
\longeq{ \dbox^2 A_{\mu}\kk{x}=0\ , }
{ \dbox B\kk{x}=0 \ . }
\ea
\ee
Those then must obey
\be
\ba{l}
  \llongeq{  \dbox{h_{{\bs k}\sigma}}^{\!\!\mu}\kk{x}=
    {f_{{\bs k}\sigma}}^{\!\!\mu}\kk{x}\ , }
    {   \dbox{f_{{\bs k}\sigma}}^{\!\!\mu}\kk{x}=0 \ , }
{ \dbox {g_{\bs k}}\kk{x} = 0 \ , }
\ea
\ee
where ${\bm k}$ and $\sigma$ denote the momentum and the polarization
of photon respectively. An explicit representation is obtained by an
orthonormal set, $\KKK{\varphi_{\bs
k}\kk{\bm p}}$,
\begin{equation}
   \sum_{\bs k}\varphi_{\bs k}\kk{\bm p}\,
   {\varphi_{\bs k}}^{\ast}\kk{\bm q}
   =\delta\kk{{\bm p}-{\bm q}}\ ,\qquad
   \inte{d^3p}\,{\varphi_{\bs k}}^{\ast}\kk{\bm p}\,
   \varphi_{\bs l}\kk{\bm p}
   =\delta_{\bs{kl}}\ ,\lab{IPP}
\end{equation}
and by a polarization vector ${\xi_\sigma}^\mu\kk{\bm p}$,
\begin{equation}
   \sum_{\sigma=0}^{3}\sum_{\tau=0}^{3}{\xi_\sigma}^\mu\kk{\bm p}
   \eta_{\sigma\tau}  {\xi_\tau}^\nu\kk{\bm p}
   =g^{\mu\nu}\ , \qquad  \sum_{\mu=0}^3 {\xi_\sigma}^\mu\kk{\bm
p}\,{\xi_\tau}_\mu\kk{\bm p}
   =\eta_{\sigma\tau}\ ,
   \lab{Axi}
\end{equation}
where ${\rm diag}(\eta_{\sigma\tau })=(1,-1,-1,-1); \ {\rm diag}( g_{\mu\nu
})=(1,-1,-1,-1)$.
With these we have
\be
\ba{l}
 \llongeq{  g_{\bs k}\kk{x}=\inte{\frac{d^3p}
     {\sqrt{\kk{2\pi}^3 2p_0}}}\,\varphi_{\bs k}
     \kk{\bm p}\,{\rmchar e}^{-ipx}\;,\mbox{\qquad}p_0=
     \abs{\bm p}\ , }
{{f_{\bs k}}_\sigma^{\;\mu}\kk{x}=
     \inte{\frac{d^3p}{\sqrt{\kk{2\pi}^3 2p_0}}}\,
     {\xi_\sigma}^\mu\kk{\bm p}\,
     \varphi_{\bs k}\kk{\bm p}\,{\rmchar e}^{-ipx}\ ,
     \mbox{\qquad}p_0=\abs{\bm p}\ , }
{  {h_{{\bs k}\sigma}}^{\!\!\mu}\kk{x}=
     {1\over2}\left({\bm \nabla}^2\right)^{-1}\KKK{\KK{x_0\der{0}{}
     -{3\over2}}{f_{{\bs k}\sigma}}^{\!\!\mu}\kk{x}
     +g^{\mu 0}{f_{{\bs k}\sigma}}^{\!\!0}\kk{x}}\ .  }
\ea
\ee
Now it is almost straightforward to see that the following relations hold:
\be
\ba{c}
 \llongeq{   \sum_{\bs k}g_{\bs k}\kk{x}\,
  {g_{\bs k}}^{\ast}\kk{y}= iD^{(+)}\kk{x-y}\ , }
{ \sum_{\bs k}\sum_{\sigma=0}^{3}\sum_{\tau=0}^{3}
  {f_{{\bs k}\sigma}}^{\!\!\mu}\kk{x}\,\eta_{\sigma\tau}\,
  {{f_{{\bs k}\tau}}^{\!\!\nu}}^{\ast}\kk{y}
  = ig^{\mu\nu}D^{(+)}\kk{x-y}\ , }
{ \sum_{\bs k}\sum_{\sigma=0}^{3}\sum_{\tau=0}^{3}
  {\overline h}_{{\bs k}\sigma}\kk{x}\,\eta_{\sigma\tau}\,
  {\overline h}_{{\bs k}\tau}^{\;\ast}\kk{y}
  =  -iE^{(+)}\kk{x-y}\ , }
 \ea
\ee
where $E^{(+)}\kk{x-y}$ and $D^{(+)}\kk{x-y}$ are the positive
frequency parts of $E\kk{x-y}$ and $D\kk{x-y}$ respectively, and
\begin{equation}
	{\overline h}_{{\bs k}\sigma}\kk{x} \equiv \der{\mu}{}\!
	      {h_{{\bs k}\sigma}}^{\!\!\mu}\kk{x} \qquad \mbox{also} \qquad
	{\overline f}_{{\bs k}\sigma}\kk{x} \equiv
	\der{\mu}{}\!{f_{{\bs k}\sigma}}^{\!\!\mu}\kk{x}\ .
%	\lab{}
\end{equation}
(There needs some regularization to define $E^{(+)}\kk{x-y}$ because
of its logarithmic divergence. However this divergence goes out when
differentiated with respect to $x_{\mu}$.)

There hold  additional relations:
\be
\ba{l}
  \longeq{   i\inte{d^3x}{{f_{{\bs k}\sigma}}^{\!\!\mu}}^{\ast}\kk{x}
     \lrder{0}{}{f_{{\bs l}\tau}}_{\mu}\kk{x}
     =\delta_{\bs{kl}}\,\eta_{\sigma\tau}\ , }
{ i\inte{d^3x}{g_{\bs k}}^{\ast}\kk{x}\lrder{0}{}
     g_{\bs l}\kk{x}=\delta_{\bs{kl}}\ . }
\ea
\ee

\subsection{Fock State and the LSZ formula}
Before constructing the LSZ formula, we define the creation and
annihilation operators,
\be
\ba{l}
 \longeq{     {{\cal A}_{{\bs k}\sigma}^{as}}^{\dag} \equiv -i\inte{d^3x}
      \KKK{{f_{{\bs k}\sigma}}^{\!\!\mu}\kk{x}\lrder{0}{}
      {A_\mu^{as}}\kk{x}+{h_{{\bs k}\sigma}}^{\!\!\mu}\kk{x}
      \lrder{0}{}\dbox{A_\mu^{as}}\kk{x}} \ ,}
      {  {\cal A}_{{\bs k}\sigma}^{as} \equiv i\inte{d^3x}
      \KKK{{{f_{{\bs k}\sigma}}^{\!\!\mu}}^{\ast}\kk{x}\lrder{0}{}
      {A_\mu^{as}}\kk{x}+{{h_{{\bs k}\sigma}}^{\!\!\mu}}^{\ast}\kk{x}
      \lrder{0}{}\dbox {A_\mu^{as}}\kk{x}} \ ,}
\ea     \lab{AOA}
\ee
\be
\ba{l}
 \longeq{  {{\cal B}_{\bs k}^{as}}^{\dag}\equiv-i\inte{d^3x}
      g_{\bs k}\kk{x}\lrder{0}{}B^{\,as}\kk{x} \ , }
   {  {\cal B}_{\bs k}^{as}\equiv i\inte{d^3x}  {g_{\bs
k}}^{\ast}\kk{x}\lrder{0}{}B^{\,as}\kk{x} \ . }
\ea     \lab{AOB}
\ee
The asymptotic one-photon states are given by
\be
\ba{l}
\longeq{  \ket{{\bm k}\sigma;as}\equiv{{\cal A}_{{\bs k}\sigma}^{as}}^{\dag}
\ket{0} \ ;
        \qquad {\cal A}_{{\bs k}\sigma}^{as}\ket{0}=0\ , }
{\ket{{\bm k};as}\equiv {{\cal B}_{\bs k}^{as}}^{\dag} \ket{0}\ ;
        \qquad \ \quad  {\cal B}_{\bs k}^{as}\ket{0}=0\ . }
 \ea       \lab{FSB}
\ee
Those satisfy, by means of \eqn{Aasycomm},
\be
\ba{l}
 \lllongeq{ \commu{{\cal A}_{{\bs k}\sigma}^{as}}{{{\cal A}_{{\bs
l}\tau}^{as}}^{\dag}}
  = -\delta_{\bs{kl}}\;\eta_{\sigma\tau}-iK\inte{d^3x}
  {\overline f}_{{\bs k}\sigma}^{\;\ast}\kk{x}\lrder{0}{}
  {\overline f}_{{\bs l}\tau}\kk{x}    }
 { \hspace{15ex} -i\kk{1-{\alpha}}\!\inte{d^3x}
  \Bigl\{{\overline f}_{{\bs k}\sigma}^{\;\ast}\kk{x}\lrder{0}{}
  {\overline h}_{{\bs l}\tau}\kk{x}+{\overline h}_{{\bs
k}\sigma}^{\;\ast}\kk{x}
  \lrder{0}{}{\overline f}_{{\bs l}\tau}\kk{x}\Bigr\}\ , }
  { \commu{{\cal A}_{{\bs k}\sigma}^{as}}{{{\cal B}_{\bs l}^{as}}^{\dag}}
  = i\inte{d^3x}{\overline f}_{{\bs k}\sigma}^{\;\ast}\kk{x}
  \lrder{0}{}g_{\bs l}\kk{x}\ , }
 { \commu{{\cal B}_{\bs k}^{as}}{{{\cal A}_{{\bs l}\tau}^{as}}^{\dag}}
   = i\inte{d^3x}{g_{\bs k}}^{\ast}\kk{x}\lrder{0}{}
  {\overline f}_{{\bs l}\tau}\kk{x}\ ,  }
  { \commu{{\cal B}_{\bs k}^{as}}{{{\cal B}_{\bs l}^{as}}^{\dag}}
  =0 \ . }
\ea  \lab{CRBB}
\ee

Armed with these we can construct the LSZ formula: to begin with, we
rewrite \eqn{AOA} as
\be
 \ba{l}
\longeq{     {{\cal A}_{{\bs k}\sigma}^{as}}^{\dag} =-i\!\inte{d^3x}
      \Bigl\{{f_{{\bs k}\sigma}}^{\!\!\mu}\kk{x}\lrder{0}{}
      {A_\mu^{as}}\kk{x}-\kk{1\!-\!{\alpha}}\!
      \bigm[\!{\overline h}_{{\bs k}\sigma}\kk{x}\lrder{0}{}
      +{f_{{\bs k}\sigma}}^{\!\!0}\kk{x}\!\bigm]\!\!B^{\,as}\kk{x}\Bigr\} \ , }
 {   {\cal A}_{{\bs k}\sigma}^{as}=i\!\inte{d^3x}
      \Bigl\{{{f_{{\bs k}\sigma}}^{\!\!\mu}}^{\ast}\kk{x}\lrder{0}{}
      {A_\mu^{as}}\kk{x}-\kk{1\!-\!{\alpha}}
      \bigm[\!{\overline h}_{{\bs k}\sigma}^{\;\ast}\kk{x}\lrder{0}{}
      +{{f_{{\bs k}\sigma}}^{\!\!0}}^{\ast}\kk{x}\!\bigm]
      \!\!B^{\,as}\kk{x}\Bigr\}\ ,  }
\ea      \lab{AOA1}
\ee
where use has been made of the equations of motion \eqn{AEq}.
Under the asymptotic condition $A_\mu \stackrel{| t | \rightarrow
\infty}{\longrightarrow} A^{as}_\mu$, we
obtain
\be
\ba{l}
\lllongeq{ {{\cal A}_{{\bs k}\sigma}^{out}}^{\dag}\Tprod{\cal O}
      -\Tprod{\cal O}{{\cal A}_{{\bs k}\sigma}^{in}}^{\dag}
      = -i\!\inte{d^4x}\Bigl\{\!{f_{{\bs k}\sigma}}^{\!\!\mu}\kk{x}
      \dbox^x\Tprod{{\cal O}{A}_\mu\kk{x}} }
{ \hspace{25ex} -\kk{1\!-\!\alpha}\!
      \bigm[\!{\overline h}_{{\bs k}\sigma}\kk{x}\dbox^x\!\!
      +\!{f_{{\bs k}\sigma}}^{\!\!\mu}\kk{x}\!\der{\mu}{x}
      \bigm]\!\Tprod{{\cal O}{B}\kk{x}}\Bigr\} \ , }
 {  {\cal A}_{{\bs k}\sigma}^{out}\Tprod{\cal O}
      -\Tprod{\cal O}{\cal A}_{{\bs k}\sigma}^{in}
     = i\!\inte{d^4x}
      \Bigl\{\!{{f_{{\bs k}\sigma}}^{\!\!\mu}}^{\ast}\kk{x}\dbox^x
      \Tprod{{\cal O}{A}_\mu\kk{x}}  }
{\hspace{25ex}  -\kk{1\!-\!{\alpha}}\!
      \bigm[\!{\overline h}_{{\bs k}\sigma}^{\;\ast}\kk{x}\dbox^x\!\!
      +\!{{f_{{\bs k}\sigma}}^{\!\!\mu}}^{\ast}\kk{x}\!\der{\mu}{x}
      \bigm]\!\Tprod{{\cal O}{B}\kk{x}}\Bigr\} \ , }
 \ea     \lab{AATTA1}
\ee
and
\be
\ba{l}
\longeq{	  {{\cal B}_{\bs k}^{out}}^{\dag}\Tprod{\cal O}
      -\Tprod{\cal O}{{\cal B}_{\bs k}^{in}}^{\dag}
      =-i\inte{d^4x}g_{\bs k}\kk{x}\dbox^x
      \Tprod{{\cal O}{B}\kk{x}} \ ,}
{  {\cal B}_{\bs k}^{out}\Tprod{\cal O}
      -\Tprod{\cal O}{\cal B}_{\bs k}^{in}
      =i\inte{d^4x}{g_{\bs k}}^{\ast}\kk{x}\dbox^x
      \Tprod{{\cal O}{B}\kk{x}} \ . }
\ea	  \lab{ABTTB}
\ee

Taking the vacuum expectation values of
\eqn{AATTA1}, we obtain the LSZ formulas:
\be
\ba{l}
\lllongeq{ \bra{0}\Tprod{\cal O}\ket{{\bm k}\sigma;in}
      = i\inte{d^4x}\Bigl\{{f_{{\bs k}\sigma}}^{\!\!\mu}\kk{x}
      \dbox^x\bra{0}\Tprod{{\cal O}{A}_\mu\kk{x}}\ket{0} }
 {\hspace{20ex} -\kk{1-{\alpha}}
      \bigm[{\overline h}_{{\bs k}\sigma}\kk{x}\dbox^x
      +{f_{{\bs k}\sigma}}^{\!\!\mu}\kk{x}\der{\mu}{x}\bigm]
      \bra{0}\Tprod{{\cal O}{B}\kk{x}}\ket{0}\Bigr\} \ , }
 {\bra{{\bm k}\sigma;out}\Tprod{\cal O}\ket{0}
    = i\inte{d^4x}
      \Bigl\{{{f_{{\bs k}\sigma}}^{\!\!\mu}}^{\ast}\kk{x}\dbox^x
      \bra{0}\Tprod{{\cal O}{A}_\mu\kk{x}}\ket{0} }
{\hspace{20ex} -\kk{1-{\alpha}}
      \bigm[{{\overline h}_{{\bs k}\sigma}}^{\ast}\kk{x}\dbox^x
      +{{f_{{\bs k}\sigma}}^{\!\!\mu}}^{\ast}\kk{x}\der{\mu}{x}\bigm]
      \bra{0}\Tprod{{\cal O}{B}\kk{x}}\ket{0}\Bigr\} \ ,}
\ea      \lab{LSZ2}
\ee
and
\be
\ba{l}
 \longeq{     \bra{0}\Tprod{\cal O}\ket{{\bm k}\,in}
      = i\inte{d^4x}g_{\bs k}\kk{x}\dbox^x
      \bra{0}\Tprod{{\cal O}{B}\kk{x}}\ket{0}  }
{\bra{{\bm k}\,out}\Tprod{\cal O}\ket{0}
      = i\inte{d^4x}{g_{\bs k}}^{\ast}\kk{x}\dbox^x
      \bra{0}\Tprod{{\cal O}{B}\kk{x}}\ket{0}  }
 \ea     \lab{LSZ4}
\ee
For the case of the Feynman gauge (${\alpha}=1$),
$\bra{0}\Tprod{{\cal O}{B}\kk{x}}\ket{0}$ terms in \eqn{LSZ2} disappear, so
that we have much simpler formulas.

\section{Invariant Regularization and Finiteness for the Energy-Momentum
Tensor}

In this appendix, we show, although might be well-known as a ``common'' sense,
that
na\"{\i}vely introduced cut-off,
\be
{ 1 \over p^2 } \longrightarrow \lim_{\Lambda \rightarrow \infty} { 1 \over p^2
}
\left({-\Lambda^2 \over  p^2 - \Lambda^2} \right)^N  \  ,
\lab{FCO}
\ee
where $N$ is some suitable number to make the whole integral finite, breaks the
Lorentz invariance but the dimensional regularization does not; since there
seems very
few examples for demonstrating this ``common'' sense explicitly. In the
subsequent
section, finiteness for the energy-momentum tensor is argued.

\subsection{A Need for Invariant Regularization}
To simplify the discussion, consider the single scalar model described by
\be
\ba{l}
\longeq{ 	{\cal L}={1\over2}\partial_{\mu}\phi\partial^{\mu}\phi
	-{\mu^{2}\over2}\phi^{2}-{\lambda\over4!}\phi^{4} }
{\hspace{3ex}
+{(Z-1)\over2}\left(\partial_{\mu}\phi\partial^{\mu}\phi-\mu^{2}\phi^{2}\right)
	-{\mu^{2}\over2}Z(Z_{\mu}-1)\phi^{2}
	-{\lambda\over4!}(Z_{\lambda}Z^{2}-1)\phi^{4}\,,  }
\ea
\ee
where all quantities are renormalized and
\begin{equation}
	\phi_{bare}=Z^{1\over2}\phi,\,\mu^{2}_{bare}=Z_{\mu}\mu^{2},\,
	\lambda_{bare}=Z_{\lambda}\lambda\,.
\end{equation}
The energy-momentum tensor is,
\begin{equation}
	\itTheta^{\mu\nu}=Z\partial^{\mu}\phi\partial^{\nu}\phi
	-g^{\mu\nu}\left({Z\over2}\partial^{\lambda}\phi\partial_{\lambda}
	-{\mu^{2}\over2}ZZ_{\mu}\phi^{2}
	-{\lambda\over4!}Z^{2}Z_{\lambda}\phi^{4}\right)\,.
\end{equation}
(For brevity's sake we do not consider an improvement; which
accuires importance in the case of the trace identity \cite{CJ}.) By following
the
standard procedure
\cite{CCJ}, the Ward-Takahashi relation (WT) for the amputated Green's
function,
$\itGamma$, is found to be
\begin{equation}
	\partial^{x}_{\mu}\itGamma^{\mu\nu}(x;y,z)
	+i\partial^{\nu}_{x}\delta^{4}(x-y)\,\itGamma(x,z)
	+i\partial^{\nu}_{x}\delta^{4}(x-z)\,\itGamma(x,y)=0\,,
	\lab{GWTT2}
\end{equation}
to give
\begin{equation}
	(k+p)_{\mu}\itGamma^{\mu\nu}(k,p)+ik^{\nu}\itGamma(-p)
	+ip^{\nu}\itGamma(k)=0\,,
	\lab{GWTT2M}
\end{equation}
where
\be
\ba{l}
\longeq{	\itGamma^{\mu\nu}(x;y,z) = \int\!\!{d^{4}k\over(2\pi)^{4}}
	{d^{4}p\over(2\pi)^{4}}{\rm e}^{-iky}{\rm e}^{-ipz}{\rm e}^{i(k+p)z}
	\itGamma^{\mu\nu}(k,p)  \   ,  }
{ \itGamma(x,y)  =   \int\!\!{d^{4}k\over(2\pi)^{4}}
	{\rm e}^{-ik(x-y)}\itGamma(k)\,. }
\ea
\ee

With the use of the Feynman cut-off \eqn{FCO}, $\itGamma(k)$ is
obtained as
\begin{equation}
	\itGamma(k)=-i(k^{2}-\mu^{2})
	-\left\{{\mit \Sigma}(k^{2})+i(Z-1)(k^{2}-\mu^{2})
	-i\mu^{2}Z(Z_{\mu}-1)\right\}\,,
\end{equation}
with
\begin{equation}
	{\mit \Sigma}(k^{2})={\mit \Sigma}(0)
	={-i\lambda\over2}\int\!\!{d^{4}l\over(2\pi)^{4}}
	{i\over l^{2}-\mu^{2}}
	\left(-\Lambda^{2}\over l^{2}-\Lambda^{2}\right)^{2}
	={-i\lambda\over2(4\pi)^{2}}\mu^{2}
	\left[{\Lambda^{2}\over\mu^{2}}
	-{\rm ln}\left({\Lambda^{2}\over\mu^{2}}\right)+1\right]\,.
\end{equation}
Put $Z=1$ and choose $Z_{\mu}$ such that
${\mit \Sigma}(0)-i\mu^{2}(Z_{\mu}-1)$ be finite.
Therefore
\be
	\itGamma(k)=-i(k^{2}-\mu^{2})
	-\left\{{\mit \Sigma}(0)-i\mu^{2}(Z_{\mu}-1)\right\}\,.
\ee
 As for $\itGamma^{\mu\nu}(k,p)$, since $Z_{\lambda}=1$ and
$Z=1$,
\be
\ba{l}
	\llongeq{ \itGamma^{\mu\nu}(k,p)  = \raisebox{-6ex}{ \epsfbox{Fig14.epsf}} }
	{  = -[k^{\mu}p^{\nu}+k^{\nu}p^{\mu}-g^{\mu\nu}kp]
	+ g^{\mu\nu}\mu^{2}Z_{\mu}
+ g^{\mu\nu}{\lambda\over2}
	\int\!\!{d^{4}l\over(2\pi)^{4}}{i\over l^{2}-\mu^{2}}
	\left(-\Lambda^{2}\over l^{2}-\Lambda^{2}\right)^{2}  }
{ + {i\lambda\over2}\int\!\!{d^{4}l\over(2\pi)^{4}}
	{l^{\mu}(l-q)^{\nu}+l^{\nu}(l-q)^{\mu}-g^{\mu\nu}l(l-q)+ g^{\mu\nu}\mu^{2}
	\over (l^{2}-\mu^{2})[(l-q)^{2}-\mu^{2}]}
	\left(-\Lambda^{2}\over l^{2}-\Lambda^{2}\right)^{2}  ,  \ }
\ea
\ee
where $q= k+p$. The integral in the first line is just the
same as ${\mit \Sigma}(0)$, then WT \eqn{GWTT2M} reads
\be
\ba{l}
\llongeq{ \lefteqn{(k+p)_{\mu}\itGamma^{\mu\nu}(k,p)+ik^{\nu}\itGamma(-p)
	+ip^{\nu}\itGamma(k)}  }
{ =  {i\lambda\over2}q_{\mu}\int\!\!{d^{4}l\over(2\pi)^{4}}
	{l^{\mu}(l-q)^{\nu}+l^{\nu}(l-q)^{\mu}-g^{\mu\nu}l(l-q)+ g^{\mu\nu}\mu^{2}
	\over (l^{2}-\mu^{2})[(l-q)^{2}-\mu^{2}]}
	\left(-\Lambda^{2}\over l^{2}-\Lambda^{2}\right)^{2}  }
{ = -{q^{\nu}\over4(4\pi)^{2}}\Lambda^{2}+O(\Lambda^{0}) \neq 0 \,.  }
\ea
\ee
Therefore WT \eqn{GWTT2M} cannot be met by the Feynman cut-off
scheme \eqn{FCO}, which implies that the translational invariance is
broken in this regularization. On the contrary adopting the  dimensional
regularization, we have, instead of \eqn{GWTT2MFC},
\be
\ba{l}
\llongeq{
\lefteqn{(k+p)_{\mu}\itGamma^{\mu\nu}(k,p)+ik^{\nu}\itGamma(-p)
	+ip^{\nu}\itGamma(k)} } { =
{i\lambda\over2}q_{\mu}\int\!\!{d^{n}l\over(2\pi)^{n}}
{l^{\mu}(l-q)^{\nu}+l^{\nu}(l-q)^{\mu}-g^{\mu\nu}l(l-q)+ g^{\mu\nu}\mu^{2}
	\over (l^{2}-\mu^{2})[(l-q)^{2}-\mu^{2}]} }
{ = {i\lambda\over2} \int\!\!{d^{n}l\over(2\pi)^{n}}
	\left\{{(l-q)^{\nu}\over(l-q)^{2}-\mu^{2}}
	- {l^{\nu}\over l^{2}-\mu^{2}}\right\} = 0 \,, }
\ea \lab{GWTT2MFC}
\ee
because of the fact that we are free to make a shift of the loop momentum
in the dimensional regularization.

\subsection{Cancellation of the Divergence in $G^{\mu\nu;\lambda\kappa}(q,q')$}

Finiteness of the energy-momentum tensor has already been proven in \cite{CCJ},
and
QED is indeed the case. To see this, we here show the cancellation of
divergences in
\be
G^{\mu\nu;\lambda\kappa}(q,q') \equiv  G^{\mu\nu;\lambda\kappa}_{\rm
g}(q,q') + G^{\mu\nu;\lambda\kappa}_{\rm m}(q,q') \ .
\ee

Up to the one-loop, in view of Fig.3, it gives
\be
\ba{l}
\llongeq{G^{\mu\nu;\lambda\kappa}(q,q')
=  { -i \over q^2} X^{\mu\nu\lambda\kappa}(q,q') {-i\over q'^2}  }
{  + { -i \over q^2} X^{\mu\nu\lambda \rho }(q,q')
{ -i \over q'^2} \itPi_{\rho\sigma}(q'){ -i \over q'^2}d^{\sigma\kappa}(q')
+ { -i \over q^2} d^{\lambda\rho}(q) \itPi_{\rho\sigma}(q)
{ -i \over q^2}X^{\mu \nu\sigma \kappa }(q,q')  { -i \over q'^2}  }
{ \hspace{0ex} + { -i \over q^2}
{d^\lambda}_\rho(q) \itPi^{\mu\nu\rho\sigma}(q, q'){d_\sigma}^\kappa(q') { -i
\over
q'^2} \ , }
\ea  \lab{G}
\ee
whose expressions from the first to the final line come from \eqn{IIfeynmrule},
\eqn{B44}, and
\eqn{II1PR} respectively.

The Lagrangian is now the Nakanishi-Lautrup one in the fully
renormalized  form,
\be
\ba{l}
\llongeq{  {\cal L} = -{1\over4}F^{\mu\nu}F_{\mu\nu}-A^{\mu}\der{\mu}{}B
  +{\alpha\over2}B^2+\bpsi({i\over2}\lrderpartb -m)\psi+e\bpsi\dslA\psi }
{ \hspace{30pt}-(Z_{3}-1){1\over4} F^{\mu\nu}F_{\mu\nu}
  +(Z_{2}-1)\bpsi({i\over2}\lrderpartb -m)\psi }
{ \hspace{30pt}-(Z_{m}-1)Z_{2}m\bpsi\psi
  +(Z_{1}-1)e\bpsi\dslA\psi\ , }
\ea
\ee
with the relation between the bare and the renormalized quantities:
\be
\ba{l}
\longeq{	A_{bare}^{\mu}=Z_{3}^{1\over2}A^{\mu} \ , \quad
	B_{bare}=Z_{3}^{-{1\over2}} B \ ,  \quad  \psi_{bare}=Z_{2}^{1\over2}\psi \ ,
}
{ \alpha_{bare}=Z_{3}\alpha\ , \quad  e_{bare}=Z_{1}Z_2^{-1} Z_3^{-1/2} e\ ,
\quad
m_{bare}=Z_{m}m\ . }
\ea
\ee
The energy-momentum tensor is therefore found as
\be
\ba{l}
\lllongeq{ \Theta^{\mu\nu}
	= -Z_{3}\left\{F^{\mu\rho}F^{\nu}_{\ \rho}
	-{g^{\mu\nu}\over4}F^{\rho\sigma}F_{\rho\sigma}\right\} }{ \hspace{5ex}
-(A^\mu
\der{}{\nu}B + A^\nu \der{}{\mu}B)
	-g^{\mu\nu}\left\{{\alpha\over2}{B}^2-A^\rho\partial_\rho B\right\}
}{ \hspace{5ex} +Z_{2}\left\{{i\over4}\bpsi
	(\gamma^\mu\!\lrder{}{\nu}+\gamma^\nu\!\lrder{}{\mu})\psi
	-g^{\mu\nu}\bpsi\left({i\over2}\lrderpartb-Z_{m}m\right)\psi\right\}
} {\hspace{5ex} +e Z_{1} \left\{
	{1\over2}\bpsi(\gamma^{\mu}A^{\nu}+\gamma^{\nu}A^{\mu})\psi
	-g^{\mu\nu}\bpsi\dslA\psi\right\} \  . }
\ea \lab{REMT}
\ee
In view of \eqn{G} divergences lies in the vacuum polarization,
$\itPi^{\mu\nu}(q)$, and $\itPi^{\mu\nu;\rho\sigma}(q, q')$.  As usual we
remove
the divergence of $\itPi^{\mu\nu}(q)$: the photon propagator up to
the one loop reads
 \be
\ba{l}
\longeq{\inte{d^{4}x}{\rmchar e}^{iqx}\bra{0}
	TA^{\mu}(x)A^{\nu}(0)\ket{0} } { \hspace{1ex} = {-i\over
q^{2} } d^{\lambda\kappa}(q)
	+{-i\over q^{2}}(g^{\lambda\kappa} q^{2}-q^{\lambda}q^{\kappa})
	\{ \itPi(q)-i(Z_{3}-1)\}{-i\over q^{2} } \ , }
\ea
\ee
where $\itPi(q)$ has been given in \eqn{IIvacpol},
\begin{equation}
	\itPi(q)=-ie^{2}{2\,{\rmchar{tr}}\mbox{\bf1}\over(4\pi)^{2}}
	\itGamma(2-{n\over2})\int^{1}_{0}\hspace{-8pt}dx\ x(1-x)
	\left(m^{2}-x(1-x)q^{2}\over4\pi\right)^{{n\over2}-2} \ .
	\lab{V}
\end{equation}
$Z_{3}$ is chosen to cancel out the divergent part of
$\itPi(q)$. While the superficial degree of
divergence for
$\itPi^{\mu\nu;\lambda\kappa}(q,q')$ is two so that the Taylor expansion in the
expression in \eqn{B47} gives
\be
\itPi^{\mu\nu;\lambda\kappa}(q,q')  = i\itPi(0)\widetilde
X^{\mu\nu\lambda\kappa}(q,q')
	+O((q,q')^{3})\ ,
\ee
where $O((q,q')^{3})$ is finite.

Now recall that \eqn{G} is rewritten in terms of the renormalized form such as
\be
\ba{l}
\llllongeq{G^{\mu\nu;\lambda\kappa}(q,q')= {-i\over q^{2}}
	Z_{3}\widetilde X^{\mu\nu\lambda\kappa}(q,q'){-i\over q'^{2}}
} { \hspace{13ex} +{-i\over q^{2}}
	\{\alpha\,q^\lambda q'^\kappa g^{\mu\nu}
	-q^\lambda X^{\mu\nu\kappa}(q,q')
	-q'^\kappa X^{\mu\nu\lambda}(q',q)
	\}{-i\over q'^{2}} }{ \hspace{13ex} +{-i\over
q^{2}}X^{\mu\nu\lambda\rho}(q,q'){-i\over q'^{2}}
	(q'^{2}g_{\rho\sigma}-q'_{\rho}q'_{\sigma})\{\itPi(q')-i(Z_{3}-1)\}
	d^{\sigma\kappa}(q'){-i\over q'^{2}} }{ \hspace{13ex} +{-i\over
q^{2}}d^{\lambda\rho}(q)
	(q^{2}g_{\rho\sigma}-q_{\rho}q_{\sigma})\{\itPi(q)-i(Z_{3}-1)\}
	{-i\over q^{2}}X^{\mu\nu\sigma\kappa}(q,q'){-i\over q'^{2}} }{
\hspace{13ex} +{-i\over q^{2}}
	\itPi^{\mu\nu;\lambda\kappa}(q,q'){-i\over q'^{2}} \ ,
}
\ea
\ee
where use has been made of the transversal condition of
$\itPi^{\mu\nu;\lambda\kappa}(q,q')$ \eqn{trv} in the final line. Note that it
has a finite
combination $\itPi(q)-iZ_{3}$; since
\begin{equation}
	\mbox{the first term + the last term} = {-i\over q^{2}}i(\itPi(0)-iZ_{3})
	\widetilde X^{\mu\nu\lambda\kappa}(q,q'){-i\over q'^{2}}
	+O((q,q')^{3}) \ .
\end{equation}
Therefore there
are no divergences in $G^{\mu\nu; \lambda\kappa}(q,q')$.

\setlength{\baselineskip}{.8\baselineskip}


\begin{thebibliography}{99}

%%%%%%%%%%%%%%%%% intro %%%%%%%%%%%%%%%%%%%%
%
\bibitem{DIRA}
P. A. M. Dirac, ``Lectures on Quantum Mechanics,'' Belfer Graduate School of
Science,  Yeshiva University, New York 1964.
%
\bibitem{BLT}
I. Bialynicki-Birula, Phys. Rev.  {\bf D2} (1970) 2877. \\
Also B.W. Lee and J.
Zinn-Justin, Phys. Rev. {\bf D5} (1972) 3121, 3137. \\
G. 't Hoot and M. Veltman, Nucl.
Phys. {\bf B50} (1972) 318.
%
\bibitem{DT}
P. A. M. Dirac, ``Principle of Quantum Mechanics,''  p. 302, Oxford University
Press, Oxford,
1958.
%
\bibitem{ST}
O. Steinmann, Ann. Phys. {\bf 157} (1984) 232.
%
\bibitem{KT}
T. Kashiwa and Y. Takahashi, ``Gauge Invariance in Quantum Electrodynamics''
(KYUSHU-HET-14, January 1994) unpublished.
%
\bibitem{LM}
M. Lavelle and D. McMullan, Phys. Rev. Lett. {\bf 71} (1993) 3758. \
Phys. Lett. {\bf 312B}  (1993) 211.
%
\bibitem{LMc}
M. Lavelle and D. McMullan, Phys. Lett. {\bf 329B} (1994) 68. Also see
``Constituent
Quarks From QCD'' (Plymouth Preprint MS-95-06).
%
\bibitem{AL}
See for example E. S. Abers and B. W. Lee, Phys. Rep. {\bf 9c} (1973) 1.
%
\bibitem{CCJ}
C. G. Callan, S. Coleman, and R. Jackiw, Ann. Phys. {\bf 59} (1970) 42.
%
\bibitem{JRK}
 J. M. Jauch and F. Rohrlich, ``The Theory of Photons and Electrons,'' p.410
$\sim$ p.415,
Spriger-Verlag, New York 1976 . \\
 T. Kashiwa, Lettere al Nuovo Cimento {\bf
16} (1976) 283,  and  Prog. Theor. Phys. (Kyoto)  {\bf 62}  (1979) 250.
%
\bibitem{CJ}
See for example, S. Coleman and R. Jackiw, Ann. Phys. {\bf 67} (1971) 552.
%
\bibitem{FW}
D. Z. Freeman, I. V. Muzinich and E. J. Weinberg,  Ann. Phys. {\bf 87} (1974)
95. \\
D. Z. Freeman and E. J. Weinberg,  Ann. Phys. {\bf 87} (1974) 354.
%
\bibitem{NN}
N. Nakanishi, Prog. Theor. Phys. {\bf 52} (1974) 1929.
%
\bibitem{NL}
N. Nakanishi,
Prog. Theor. Phys. {\bf 35} (1966) 1111; {\bf 49} (1973) 640.\\
B. Lautrup, Mat. Fys. Medd. Dan. Vid. Selsk. {\bf 35} (1967) 29.
%
\bibitem{BRS}
C. Becchi,A. Rouet and R. Stora, Ann. Phys. {\bf 98} (1976) 287. \\
T. Kugo and I. Ojima, Phys. Lett. {\bf 73B} (1978) 459.
%
\bibitem{KO}
T. Kugo and I. Ojima, Prog. Theor. Phys. Suppl.  No. 66,  (1979) 1.
%
\bibitem{BD}
J. D. Bjorken and S. D. Drell, ``Relativistic Quantum Fields,'' p.197,
McGraw-Hill, Inc. 1965.
%
\bibitem{PR}
P. G. Federbush and K. A. Johnson, Phys. Rev. {\bf 120} (1960) 1926. \\
P. Roman, ``Introduction to Quantum Field Theory,'' p.381, John Wiley \& Sons,
Inc. 1969.
%


%%%%%%%%%%%%% Sec. 3 %%%%%%%%%%%%%%%%%%%%%%%%%%
%
\bibitem{GA}
H. D. I. Abarbanel and J. Bartels, Nucl. Phys. {\bf 136} (1978) 237 .\\
V. N. Gribov, Nucl. Phys. {bf 139} (1978) 1.
%

%%%%%%%%%%%%%%% Sec.4 %%%%%%%%%%%%%%%%%%%%%%%%%%%
\bibitem{JSCL}
J. L. Gervais and B. Sakita, Phys. Rev. {\bf D18} (1978) 453. \\
N. H. Christ and T. D. Lee, Phys. Rev. {\bf D22} (1980) 939.
%
\bibitem{FJ}
R. Floreanini and R. Jackiw, Phys. Rev. {\bf 37} (1988) 2206.
%
\bibitem{KAS}
T. Kashiwa, Prog. Theor. Phys. {\bf 70} (1983) 1124.
%
\bibitem{KS}
T. Kashiwa and M. Sakamoto, Prog. Theor. Phys. {\bf 67} (1982) 1927. \\
Also see T. Kashiwa, Prog. Theor. Phys. {\bf 66 } (1981)  1858.
%
\bibitem{Tab}
Y. Takahashi, Physica, {\bf 31} (1965) 205.
%
\bibitem{FS}
L. D. Faddeev and A. A. Slavnov, ``Gauge Fields,'' chap.3,  Benjamin, Inc.
1980.
%
%%%%%%%%%%%%%%%%%% sec.5 %%%%%%%%%%%%%%%%%%%%%%%%
%
\bibitem{NP}
P. Nelson and L. Alvarez-Gaume,  Comm. Math. Phys. {\bf 99} (1985) 103.
%
\bibitem{JRa}
 J. M. Jauch and F. Rohrlich, ``The Theory of Photons and Electrons,''   p.485,
Spriger-Verlag, New York 1976 .
%
\bibitem{WC}
K. G. Wilson,  Phys. Rev. {\bf D10} (1974) 2445 . \\
M. Creutz, ``Quarks Gluons and Lattices,''  Cambridge University Press 1983.
%
\bibitem{TO}
Y. Ohnuki and S. Kitakado, J. Math. Phys. {\bf 34} (1993) 2827. \\
D. McMullan and I. Tsutsui, Ann. Phys. {\bf 237} (1995) 269.
%
\bibitem{ka}
T. Kashiwa, Prog. Theor. Phys. {\bf 95} (1996) 421.
%
%%%%%%%%%%%%%%% Appendix %%%%%%%%%%%%%%%%%%%%%%%%
\bibitem{NoNa}
N. Nakanishi, Prog. Theor. Phys. Suppl. {\bf 51} (1972) 1.
%
\bibitem{GB}
S. N. Gupta, Proc. Phys. Soc. {\bf A63} (1950) 681. \\
K. Bleuler, Helv. Phys. Acta {\bf 23} (1950) 567.
%
\end{thebibliography}
\end{document}